\documentclass[reprint, amsmath, amssymb, aps, pre]{revtex4-1}
\usepackage{natbib}
\usepackage{graphicx}
\usepackage{dcolumn}
\usepackage{bm}
\usepackage{color}
\usepackage{mathtools}

\usepackage{hyperref}

\begin{document}

\preprint{APS/123-QED}

\title{Mesoscale computational protocols for the design of highly cooperative bivalent macromolecules}  

\author{Suman Saurabh}
\email{suman.saurabh@cnrs-orleans.fr}
\affiliation{Centre de Biophysique Moleculaire, CNRS, Rue Charles Sadron, 45071, Orl\'eans, France.}%
\author{Francesco Piazza}%
\affiliation{Universit\'e d'Orl\'eans, P\^ole de Physique, 45071, Orl\'eans, France.}%
\affiliation{Centre de Biophysique Moleculaire, CNRS, Rue Charles Sadron, 45071, Orl\'eans, France.}%

\date{\today}

\begin{abstract}
\noindent The last decade has witnessed a swiftly increasing interest in the design and production 
of novel multivalent molecules as powerful alternatives for 
conventional antibodies in the fight against cancer and infectious diseases. 
However, while it is widely accepted that large-scale flexibility ($10-100$ nm) and free/constrained 
dynamics (100 ns $- \mu$s) control the activity of such novel molecules, 
computational strategies at the mesoscale still lag behind experiments in optimizing
the design of crucial features, such as the binding cooperativity (a.k.a. avidity).\\
\indent In this study, we introduced different coarse-grained models of a polymer-linked, 
two-nanobody composite molecule, with the aim of laying down the physical bases of a thorough 
computational drug design protocol at the mesoscale. 
We show that the calculation of suitable potentials of mean force allows one 
to apprehend the nature, range and strength of the thermodynamic forces 
that govern the motion of free and wall-tethered molecules. Furthermore, we develop a 
simple computational strategy to quantify the encounter/dissociation dynamics 
between the free end of a wall-tethered molecule and the surface, at the roots of binding cooperativity. 
This procedure allows one to pinpoint the role of internal flexibility and weak non-specific interactions 
on the kinetic constants of the NB-wall encounter and dissociation.
Finally, we quantify the role and weight of rare events, which are expected to play a major
role in real-life situations, such as in the immune synapse, 
where the binding kinetics is likely dominated by fluctuations.    
\end{abstract}

\maketitle

%
%

\section*{\label{sec:intro}Introduction}
\noindent Single-domain antibodies, also known as {\em nanobodies} (NB)~\cite{nanobodies}, 
are found naturally in camelids and represent an intriguing alternative to design and build
novel multivalent and multi-specific immunotherapy agents 
for targeting tumors and viral infections~\cite{int-1}. 
In addition to leading to an increase in affinity, coupling two (or more) different {\em binders} 
helps in marginalising the effect of mutation and polymorphism of the target. 
Linked anti-CD16 nanobodies, C21 and C28~\cite{anti-cd16}, for example, have been shown to be 
effective against treating breast cancer involving a low HER2 expression, 
which is resistant to the therapeutic antibody trastuzumab~\cite{int-2}. \\
\indent As the conventional antibodies are large structures, they may not be effective when 
it comes to situations like a partially exposed tumor~\cite{tumor-0, tumor-1}. In addition, an uneven 
distribution of the anti-tumor antibody in the entire tumor region may lead to  tumor regrowth~\cite{tumor-2}. 
In such situations, therapeutic agents with smaller structures are desirable~\cite{small}. 
A good solution is to use engineered structures composed of antigen-specific nanobodies linked with flexible 
linkers~\cite{dbd-1}, e.g. realized via a polymer such as [Gly$_4$Ser]$_n$. 
Such structures will be small and will thus have 
better ability to penetrate into the tumor microenvironment and reinforce the formation of the immune synapse. 
Such structures are easy to produce and are more efficient when compared to the bulky conventional antibodies, 
and can be designed to be multivalent and multispecific~\cite{eprot-1, eprot-2}.\\
\indent An important property of an antibody is the strength of bivalent binding 
that it demonstrates, a property known in immunology as {\em avidity}. Although it has no unique quantitative 
definition (let alone whether it is more appropriate to regard it as an equilibrium or a kinetic
parameter), avidity can be thought of as the cooperative gain in affinity afforded by double binding.
Notably, such measure is intimately connected to the internal flexibility of the multi-domain molecules and 
to the geometric configuration of binding epitopes~\cite{Abpiazza}.
For example, virions employ various strategies to evade the action of antibodies of the immune system. 
Some virions, like HIV-1, have a very rapid rate of mutation. Experiments reveal that the enhanced antibody 
evasion capability of HIV-1 is based not just in its capability to mutate but is a combined effect of mutation 
along with the spike structure and low spike density~\cite{eva-1, eva-2, eva-3}. While, mutations reduce the 
affinity of the natural antibodies towards the target spikes, the spike structure and low density preclude intra- 
(i.e. multiple epitopes on the same target) 
and inter-spike cross-linking  thus preventing bivalent binding and avidity~\cite{eva-4, eva-5}.\\
\indent Experiments have demonstrated that linking two Fab domains of an antibody 
through an extended polymer like a DNA oligomer, leads to an increase in divalent binding, 
for an optimal linker length, and a resulting increase in the efficacy of the antibodies. 
Wu~\textit{et al.}~\cite{Wu} performed experiments on respiratory syncytial virus, which has a very 
high Env spike density, and showed that the affinity of low-affinity bivalent Fabs was 2$-$3 orders of 
magnitude higher as compared to their monovalent counterparts and the efficiency was not affected by 
mutations that increased the off-rates nearly 100-fold.
The results show that multivalent structures made of polymer linked nanobodies would lead to higher degree of 
avidity and thus higher efficiency. 
Galimidi~\textit{et al.} \cite{galimidi} performed experiments on linked Env (HIV-1 envelope glycoproteins) 
binders and showed that linking can lead to an increase in potency by 2$-$3 orders of magnitude. 
J\"{a}hnichen et al.~\cite{jahnichen} developed two different single domain nanobodies that could bind to different 
sites on the extracellular domain of the CXCR4 coreceptor. They found that joining the two nanobodies with protein 
linkers resulted in a 27-fold increase in CXCR4 affinity. With further analysis they concluded that the 
effect is pure avidity resulting from the heterobivalent linking of the two nanobodies. 
Zhang~\textit{et al.}~\cite{zhang} developed multimers of nanobodies leading to an increase in affinity and 
several orders of magnitude decrease in the rates of dissociation. Yang~\textit{et al.}~\cite{yang} 
performed dissociation rate calculations for the binding between a bivalent antibody and hapten ligands as a 
function of the ligand density. They found cooperative binding as the hapten density increased and bivalent binding set in. 
They could determine two different dissociation constants for the double-step antibody-hapten bnding process 
with one dissociation constant being 3-orders of magnitude larger than the other.\\
\indent When one of the nanobodies in a two-NB construct binds to the receptor at its binding site (epitope), 
the other linked ligand spends more time in the vicinity, leading to a larger probability of the 
latter unit to bind to another similar or different epitope, on the same or on a facing surface, 
depending on whether the system is mono-specific or multi-specific. 
By hindering free diffusion of the ligands, linking can lead to an increase in rebinding events 
and strengthen the interaction between interfaces~\cite{fasting}. Further, Bongrand~\textit{et al.}~\cite{bongrand} 
showed that the fraction of divalent attachments between an antibody-coated microsphere and a mono- or divalent 
ligand-coated surface, that resisted a force of 30 pN for a minimum of 5 seconds, 
was $\sim$~4 times higher than the number of monovalent attachments.\\
\indent While the choice of the nanobody depends not only on its affinity towards the target epitope, but also on 
the nature of bond it forms with the epitope~\cite{nk}, the choice of linker would depend on its flexibility and 
the geometry of the epitope distribution/configurations. 
Experimental methods have been developed to create linkers of given stiffness and extension, 
out of a combination of proteins and peptides~\cite{lin-1, lin-2, lin-3, lin-4}. 
Among the most common are the (Gly$_4$Ser)$_n$ linkers. Protein linkers have been accommodated 
into the hinge regions of natural antibodies, thus enabling intra-spike linking to viral receptors~\cite{lin-3}. 
Other biocompatible polymers like PEG are also good candidates to be used to link the nanobodies. 
The properties of the linker are very important in determining the degree of avidity. 
Depending on the epitope density on a tumor cell or the distribution of the antibody binding sites on the viral envelope, 
a linker that is either too flexible or too stiff  can lead to under-performance.\\
\indent With the improvements in computational resources and speed, molecular dynamics (MD) simulations
in recent days have been playing an important role in fields like drug discovery~\cite{drug-design} and also 
in unraveling fundamental mechanisms involving very large biological complexes, such as chromatin~\cite{billion}. 
MD simulations can be important in determining the optimal properties of the linkers that would lead to an 
efficient multivalent binding for a particular target. In addition, simulations of a group of 
linked nanobodies can give important insights into their epitope-binding kinetics as a function 
of the linker structural properties and other important parameters, like the paratope-epitope binding energy. 
While the engineered nanobody-linker-nanobody systems are much smaller as compared to the conventional antibodies, 
simulating a significantly large group of them in atomistic detail would be computationally expensive and cannot be 
done routinely. Thus, some degree of coarse-graining is important to perform kinetics analysis using MD 
simulation as a tool. \\
\indent Keeping the above discussion in mind, here we perform MD simulation of a coarse-grained system 
consisting of two nanobodies connected by a linker (referred to as a diabody) and study its structural and 
dynamical properties. We use different levels of coarse graining schemes to represent the diabody. 
In the simplest representation, we perform simulation of two extended (rigid) spheres connected by a 
flexible bead-string linker 
(see Fig.~\ref{FIG.modelsph}). In addition to this, with a future aim to study the dynamics of diabodies in the
presence of their target receptors (such as HER2), where a nanobody represented by a hard sphere will be incapable of 
representing the important features of the diabody-target interaction properly, 
we perform a finer coarse-graining of the nanobody (see Fig.~\ref{FIG.modelsph}). 
In this scheme, we represent the nanobody using the shape-based coarse-graining (SBCG) 
scheme developed by Schulten~\textit{et al.}~\cite{sbcg}. Using Umbrella Sampling (US) simulations 
we calculate the free energy profiles on which various conformations of diabodies tethered to a surface lie. 
From the MD trajectories we calculate the flight and residence times of the free end of the tethered diabody 
in a region close to the tethering wall. Comparing the results from the two different models we demonstrate how the coarse-graining scheme 
could affect the results and, notably, we highlight the role of the "shape" in the wall-domain dynamics. \\
\indent The paper is organize as follows. In section~\ref{sec:methods} we provide the details of our 
computational methods, models and simulations. In  section~\ref{sec:results}, we describe and comment 
our results, mainly concerning the different potentials of mean force and the detailed analyses of the
encounter and dissociation kinetics of the free NB of a wall-tethered molecule. In   section~\ref{sec:concl}
we wrap up our discussion and provide some final comments on this work and its perspectives.

\section{\label{sec:methods}Methods}

\subsection{Coarse-graining schemes and simulations}

\begin{figure}
\centering
\includegraphics[width=\columnwidth]{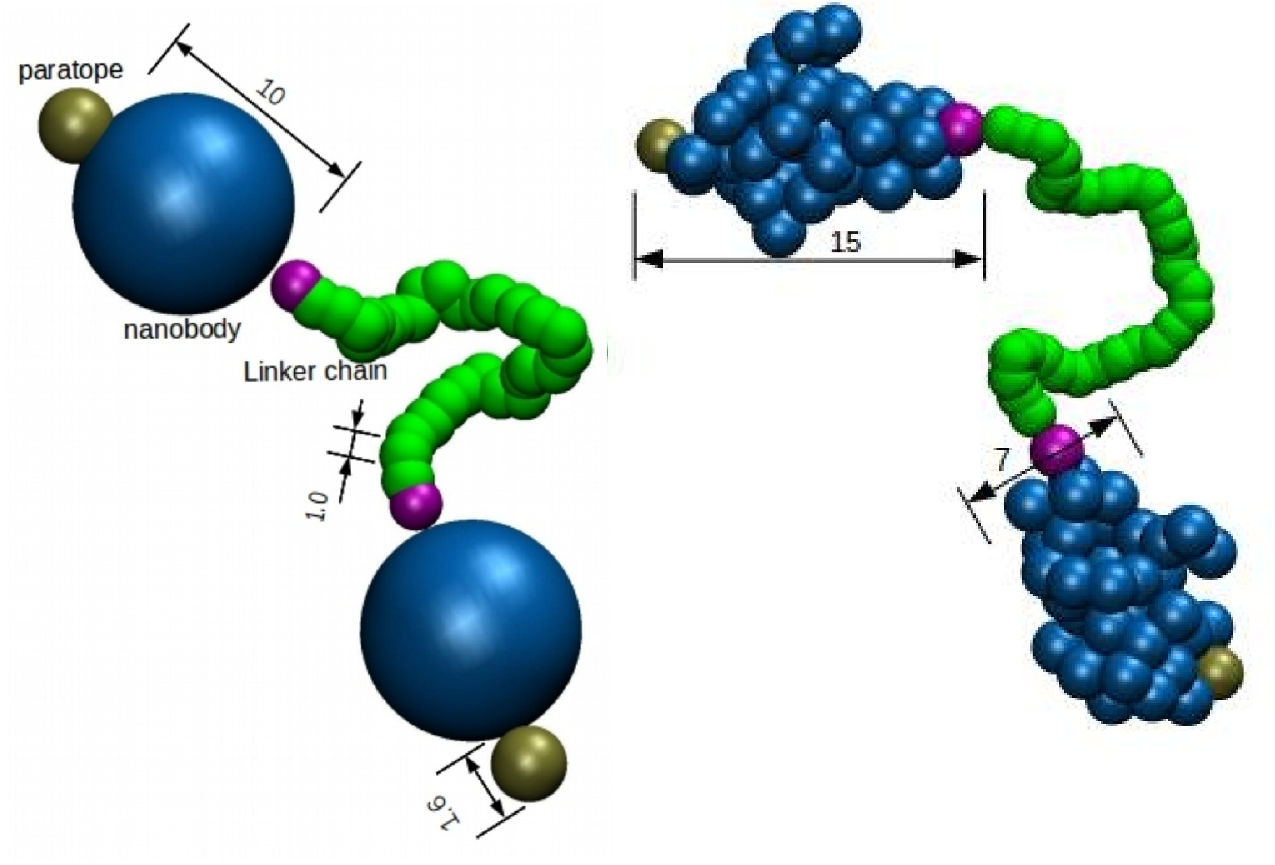}
\caption{(Color online) The two coarse-grained models of diabody studied in this work. 
Two nanobodies are linked through a polymer. The nanobody beads are represented in blue an the polymer in green. 
The maroon beads are the connector beads, which connect the nanobody to the polymer, and paratopes, respectively. 
The model on the left (named SPH) uses single spherical beads to represent the nanobodies. 
The model on the right (named SBCG) uses a shape-based coarse-graining algorithm for the nanobodies.}
\label{FIG.modelsph}
\end{figure}

\noindent All the data reported in this work were generated from Langevin dynamics simulations 
performed with LAMMPS~\cite{lammps} in the Lennard-Jones (LJ) unit system, with the unit 
of length being $\sigma$ = 3.5 \AA \ (the size of one monomer of the linker) 
and the unit of energy being $\epsilon$ = 100 K. 
All the simulations were conducted via Langevin dynamics in the overdamped regime.\\   
\indent In this work, we considered two different coarse-grained representations of a two-NB molecule 
joined by a flexible linker. In the first approach, the NBs were modeled as rigid spheres of radius 10 
(in units of linker monomers), decorated with two smaller fixed spheres at two diametrically opposite ends, 
representing the NB-linker connecting unit and the paratope, respectively   (see Fig.\ref{FIG.modelsph}).
The paratope bead has diameter 1.6 and the connector bead diameter 1. 
A bead-spring polymer bridges the two connector beads. The beads constituting the polymer linker 
have unit diameter. This model will be referred to in the following  as the SPH model.\\
%
%
\begin{figure*}
\centering
\includegraphics[width=1.5\columnwidth]{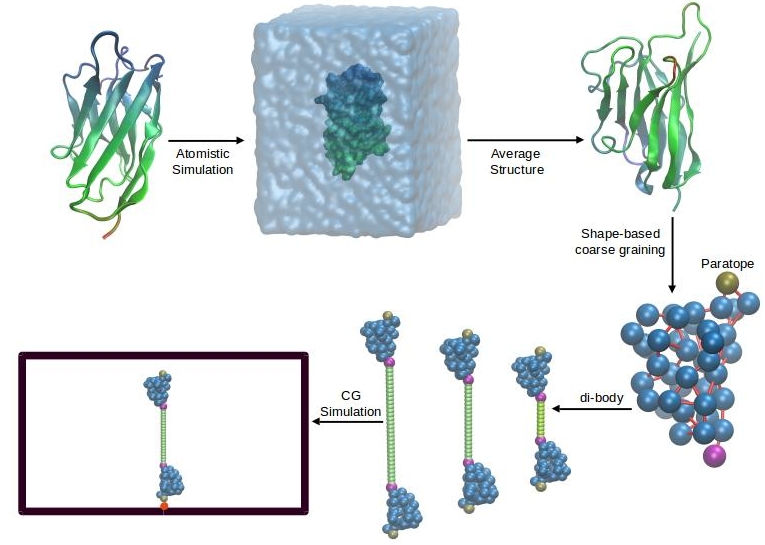}
\caption{(Color online) \textbf{The coarse-graining procedure}. The crystal structure (PDB id. 1qd0) is used as the starting 
configuration for the explicit solvent atomistic simulation. Snapshots from the last 10 ns 
of the trajectory are used to generate an average structure. The average structure is then coarse-grained 
using the shape-based coarse-graining method. A pair of the coarse-grained structures are connected 
by a linker (10-, 20- and 30-mer), enclosed in a box with $X$-$Y$ periodicity, 
aligned along the $Z$-direction and tethered to the box base, thus generating the starting 
configuration for CG simulations.}
\label{FIG.cg}
\end{figure*}
%
The second model was meant to reproduce the {\em shape} and large-scale flexibility of the NBs. 
For that we used the shape-based coarse-graining scheme developed in Schulten's group~\cite{sbcg}. 
This procedure requires a trajectory from an atomistic equilibrium MD simulation to be sampled and fed as an input.    
The crystal structure with pdb id: 1qd0 \cite{1qd0} was used as the starting structure 
for the MD simulation to generate the input structure. 
This is a camelid heavy chain variable (VHH) domain, in complex with a RR6 dye dimer.\\
\indent The dye was removed from the complex and the remaining protein was solvated in TIP3P water with a 20 \AA\ buffer,
leading to a system size of 40298 atoms, with 13284 water molecules. 
The solvated system was then neutralized by adding 5 Cl$^-$ ions to generate the starting 
configuration for the MD simulation (see Fig.\ref{FIG.cg}).
The system was minimized for 10000 steps using the conjugate gradient method. 
During minimization, all the atoms in the protein were constrained to their starting positions. 
This allowed water molecules to re-organize and eliminate unfavorable contacts with the protein. 
After minimization, the atoms were assigned velocities generated from a Maxwell distribution at 300 K. 
The particle mesh Ewald (PME) method with a real space cut-off of 12 \AA\ was used to estimate the energy 
component from the long-range electrostatic interaction. The system was simulated for 100 ns in the NPT ensemble.  
A Langevin thermostat was used~\cite{langevin} with a temperature coupling constant of 5 ps$^{-1}$, while the 
pressure was regulated using the Langevin barostat with a pressure coupling constant of 50 ps$^{-1}$. 
The simulation was performed using NAMD~\cite{namd} and the CHARMM 27~\cite{charmm} force field was used to 
describe the protein. The MD trajectories were visualized using VMD \cite{vmd}.\\
\indent An average structure was generated from the snapshots belonging to the last 10 ns of the MD 
trajectory (see Fig.~\ref{FIG.cg}). The average structure of the nanobody was used to generate a shape-based 
coarse grained (SBCG) model using the procedure formulated by Schulten \textit{et al.}~\cite{sbcg}. 
This scheme uses topology-conserving algorithm developed for neural networks to generate a 
coarse-grained representation that reproduces the shape of the protein. In a trade-off between the system size 
and a good representation of the protein shape, we used 40 beads to represent the 126 residue protein. 
To generate the system to be simulated two of the CG proteins were connected by a polymer linker with 
monomer diameter 1, similar to the one used in the SPH model (see Fig.\ref{FIG.modelsph}). 
It is to be noted that the diameter of the spherical bead that represents the nanobody in the SPH model 
is nearly equal to the geometric average of the major and minor axes of the roughly spheroidal SBCG nanobody. 
In this sense the two models are equivalent and comparable.

\subsection{Interaction parameters}
\noindent In general, the total interaction potential of our coarse-grained models 
had bond, angle and van der Waals (vdW) terms, given by
\begin{eqnarray}
&&V_{ij}^{bond} = \frac{1}{2}k_b(r_{ij}-r_0)^2 \label{e:kb} \\
&&V_{ijk}^{angle} = \frac{1}{2}k_\theta(\theta_{ijk}-\theta_0)^2 \label{e:ktheta}\\
&&V_{ij}^{vdW} = 4\epsilon_{ij}\left[ 
                                   \left( \frac{s_{ij}}{r_{ij}} \right)^{12}
                                 - \left( \frac{s_{ij}}{r_{ij}} \right)^6
                               \right] \label{e:vdW}
\end{eqnarray}
where $k_b$ and $k_\theta$ are the bond force constant and the angle bending energy, respectively, 
$r_0$ and $\theta_0$ are the equilibrium bond length and angle, respectively, $\epsilon_{ij}$ is the LJ 
interaction energy and $s$ is the inter-bead distance at which the LJ 
potential becomes repulsive  (referred to as {\em repulsive} length in the following), 
which depends on the combined radii of the interacting beads.\\
%

\subsubsection{Interaction parameters for the SBCG nanobodies and the linker}

\noindent All the CG beads were kept neutral. The connectivity and spring constants for the bonds between 
the beads were used as generated by the SBCG scheme. It is to be noted that the bonds are not set 
by a distance-based cut-off scheme, but are in accordance with the bonds present in the atomistic system. 
This helps in maintaining the flexibility of the nanobody and providing the required flexibility to the loop regions, 
which may play a defining role in the kinetics of the diabody. 
Repulsive vdW interactions among 
the beads were also introduced to ensure that the shape is maintained during the course of the simulation. 
The vdW radii of the protein beads were generated by comparing the masses of the protein beads to that of 
the PEG monomer. The repulsion between the constituent beads of the nanobody was represented by a 
Weeks-Chandler-Andersen potential (WCA)~\cite{WCA}, i.e. a shifted
LJ potential cut off at the minimum, i.e. r$_{cut}$ = 2$^{1/6}\sigma$.
The angle parameters for the nanobody beads were used as generated by the SBCG scheme.\\
\indent The linker is represented as a freely-jointed chain with two-body harmonic bond-stretching
and three-body harmonic bond-bending potentials.
The masses and diameters of the polymer beads were set to 1. The vdW interaction between 
the bead pairs was set to purely repulsive as represented by a WCA potential.
The force constant of the (stiff) bond-stretching potential was set to 54 N/m, which corresponds to
an average fluctuation of the bonds at room temperature of about 2 \% of the equilibrium length. 
In order to estimate the appropriate value of the bending rigidity $k_\theta$, 
it is expedient to refer to the calculation of the persistence length $\ell_p$
for the freely jointed chain with angle-bending interactions in the 
hypothesis of zero correlation between bending and torsion degrees of freedom.
Referring to published data for PEG~\cite{params}, we fixed $k_\theta = 1.8 \, k_BT$
as the bending coefficient for the linker
(see~\cite{suppl} for a detailed discussion).

\subsubsection{Interaction of the beads with the box walls}

\noindent The walls of the simulation box in the $X$ and $Y$ directions had periodic boundary condition, 
while the $Z$ walls were fixed. The Z walls interact with the beads via LJ (12-6) interaction given by
\begin{equation}
V_{wall}^{vdW} = 4\epsilon   \left[ 
                                     \left( \frac{s_{wall}}{r_{wall}} \right)^{12}
                                    -\left( \frac{s_{wall}}{r_{wall}} \right)^6
                             \right]
\end{equation}
Here $r_{wall}$ is the distance of the center of any bead from the wall. The wall interaction parameters, $s_{wall}$, $r_{cutwall}$ and $\epsilon$ 
define the nature of interaction between different beads and the wall. 
For the purely repulsive wall, $r_{cutwall}$ = 2$^{1/6}s_{wall}$, 
while for the attractive wall it equals $2.5\,s_{wall}$ (see also Table~\ref{tab1}). 
While the interaction of the nanobodies and paratopes with the 
$z$ walls was set to be either repulsive or attractive in different simulations, 
the linker and connector beads always had repulsive interaction with the walls.

\subsection{Preparing the systems for simulation}

\begin{figure}[t!]
\centering
\includegraphics[width=0.5\columnwidth]{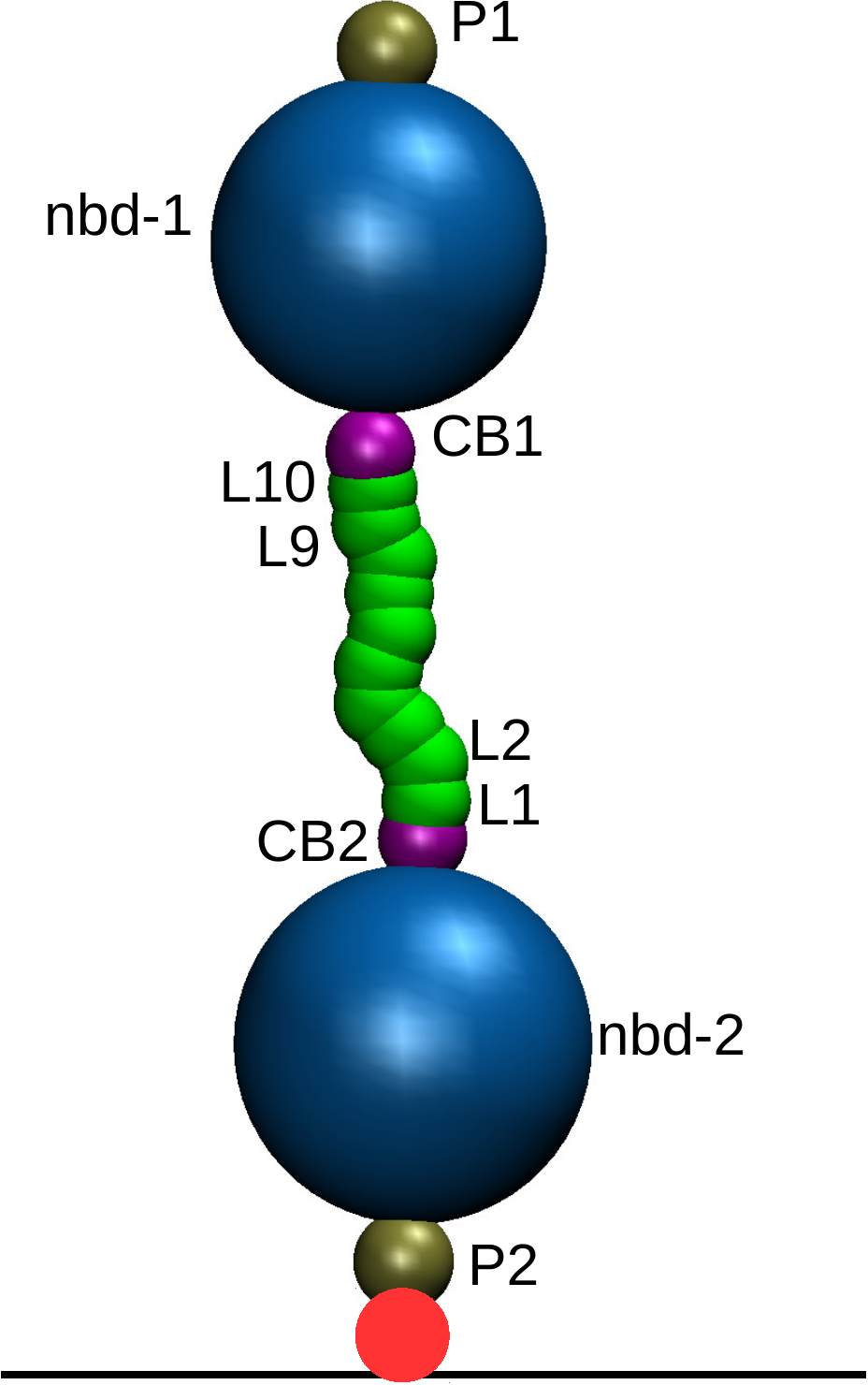}
\caption{(Color online) A 10-mer SPH diabody with the labels of different beads. 
The same naming scheme is used for the SBCG diabody. 
The tethering point is represented by the red bead.}  
\label{FIG.bead-name}
\end{figure}

\noindent The first set of simulations reported involve the calculation of potentials of mean force (PMF) 
for the diabodies as a function of various reaction coordinates. To perform the PMF calculations, 
the diabodies were enclosed in cuboidal boxes which were periodic in the $X$ and $Y$ directions, while the $Z$ 
walls were fixed and repulsive. The diabodies were tethered to the lower $Z$ wall by imposing an attractive LJ 
interaction between the paratope of one of the nanobodies and a fixed epitope bead attached to the lower $z$-wall. 
The simulations were performed for linker lengths of 10, 20, 30, 40 and 50 monomers
for the SPH system and 10, 20 and 30 monomers for the SBCG system. \\
\indent The second kind of simulations were performed on a collection of diabodies to calculate dynamical parameters. 
$N = 25$ diabodies were tethered to the lower $Z$-wall of a cuboidal box. The tethering points were arranged in a 
$5 \times 5$  lattice (see Fig.~2 of supplementary information (SI)~\cite{suppl}). 
Again, the $x$ and $y$ directions had periodic boundary conditions, while the lower $z$ wall was fixed 
and either perfectly reflecting or attractive. 
The linker lengths for different systems were similar to that considered for the PMF calculation, with an additional linker length of 60-mer for the SPH system.\\
\indent In the rest of the article, the free nanobody is referred to as nbd-1 while the nanobody tethered to the wall 
is referred to as nbd-2. The connector bead corresponding to nbd1 and nbd2 are named CB1 and CB2 respectively, 
while the paratopes are referred to as P1 and P2 respectively (see Fig.\ref{FIG.bead-name}).
In addition to the various bonded and non-bonded interactions described in the previous section, the angles 
L10-CB1-P1, P2-CB2-L1, L2-L1-CB2 and L9-L10-CB1 were restrained to 180$^{\rm o}$ using a harmonic angle bending potential 
of the form of eq.~\eqref{e:ktheta}
with $\theta_0$ = 180$^{\rm o}$ employing stiffer bending coefficients as compared to the angles corresponding to the linker. On top of this, for the SBCG systems, the nanobodies were restrained 
from rotating about their respective long axes by restraining a dihedral formed by L10-CB1 (L1-CB2) and two beads of 
nbd-1 (nbd-2) to their starting values throughout the simulation. The extra restraints were introduced to mimic the 
fact that, in real-life systems, the bonds between the linker and the nanobody 
restrict the angular motion of the nanobody about the CB1/2-P1/2 axis.

\subsection{PMF calculation} 

\begin{figure}
\centering
\includegraphics[width=\columnwidth]{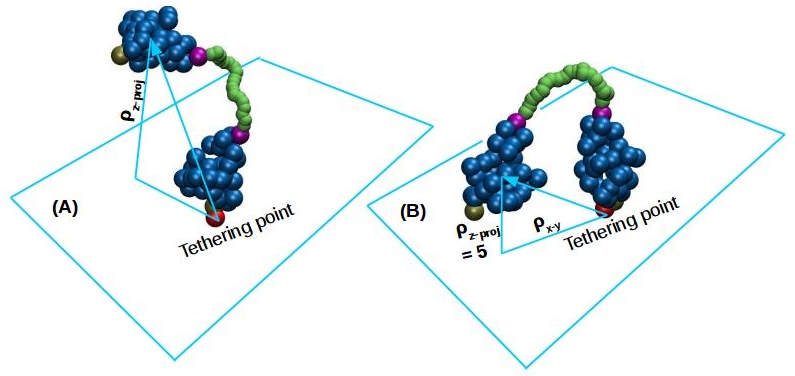}
\caption{(Color online)  \textbf{Reaction coordinates}. Schematic representation of 
(A) $\rho_{z-proj}$ and (B) $\rho_{x-y}$. The red bead is the tethering point.}
\label{FIG.rc}
\end{figure}

The PMF calculations have been performed using the umbrella sampling (US) technique along
two different reaction coordinates (RCs). The two RCs are named $\rho_{z-proj}$ and $\rho_{x-y}$. 
The former is the projection of the vector joining the tethering point to the center of mass of nbd-1 on 
the $z$-axis (see Fig.\ref{FIG.rc} (A)). The latter is the projection of the same vector on the $x-y$ 
plane with the condition that $\rho_{z-proj}$ = 5$\,\sigma$, which represents a condition where nbd-1 
is close to the wall to which nbd-2 is tethered (see Fig.\ref{FIG.rc} (B)).
%
%
The RC values varied from 3 to 30, 40, 50, 60 and 70$\,\sigma$ for the 10-, 20- and 30-, 40- and 50-mer linker systems, 
respectively, with windows at gaps of $0.5\,\sigma$, leading to 55, 75, 95, 115 and 135 
simulation windows for the SPH systems. 
For the SBCG systems, the RC values varied from 5 to 35, 45 and 55$\,\sigma$ for the 10-, 20- and 30-mer linker systems, 
leading to 61, 81 and 101 windows. The US simulations in different windows were performed in parallel. 
To generate the starting structure for each window, a short simulation was performed with the free nanobody being 
dragged from the initial to the final value of the reaction coordinate with an equilibration time of $5 \times 10^5$ steps 
at each value, and the final snapshots at each value were used as the starting structures for different windows. 
Starting from the structures thus generated, simulations lasting for $5 \times 10^7$ time steps were performed 
in each US window.

\subsection{Calculation of flight/residence times}

\noindent The different kinds of simulations performed to calculate the flight and residence times are listed in Table I. 
We simulated a system consisting of a group of 25 diabodies arranged in a $5 \times 5$ square lattice, tethered to 
the lower $Z$  surface (see~\cite{suppl}). The distance between two neighboring diabodies was set such 
that no interaction would be possible between them at any time. 
An initial simulation of 15 $\mu$s was performed. The final structure was used to 
perform three independent simulations lasting for 30 $\mu$s for the SPH system (linker lengths: 10$-$60). 
For the SBCG system (linker lengths: 10$-$30) one single simulation lasting 15 $\mu$s was performed for each linker 
length for comparison. In these simulations, the $Z$-walls were repulsive. Additional simulations were performed 
for the SPH system for all different linker lengths, with slightly attractive $z$-walls. The value of 
$\epsilon$ for these simulations was set to 1.5 $k_BT$ and 2.5 $k_BT$ for two different sets of simulations.
The $z$-coordinate of P1 was recorded and a threshold $z_{th} = 3.5 \,\sigma$ between P1 and the tethering wall was set to 
distinguish between flight and residence. More precisely, a series of consecutive simulation frames during which P1 
stayed below the threshold was considered to be a residence event, while flight events (and corresponding times) 
were associated with consecutive  frames where the paratope remained above the threshold. 
Taking an average over the 25 nanobodies (nbd-1) and three independent simulations, 
we computed the average flight and residence times and also calculated the corresponding distributions.

\section{\label{sec:results}Results and discussion}

\noindent We performed a 10 ns long simulation of the SBCG nanobody and used the trajectory to 
measure the radius of gyration of the coarse-grain model. We found the average over the trajectory to 
be $3.99 \pm 0.06 \, \sigma$. The same calculation over the last 10 ns of the atomistic trajectory 
of the \texttt{1qd0} structure yielded an average value of $4.12 \pm 0.02 \,\sigma$, which confirmed the 
soundness of our SBCG-based approach.

\subsection{Free energy profiles}

\begin{figure}
\centering
\includegraphics[width=\columnwidth]{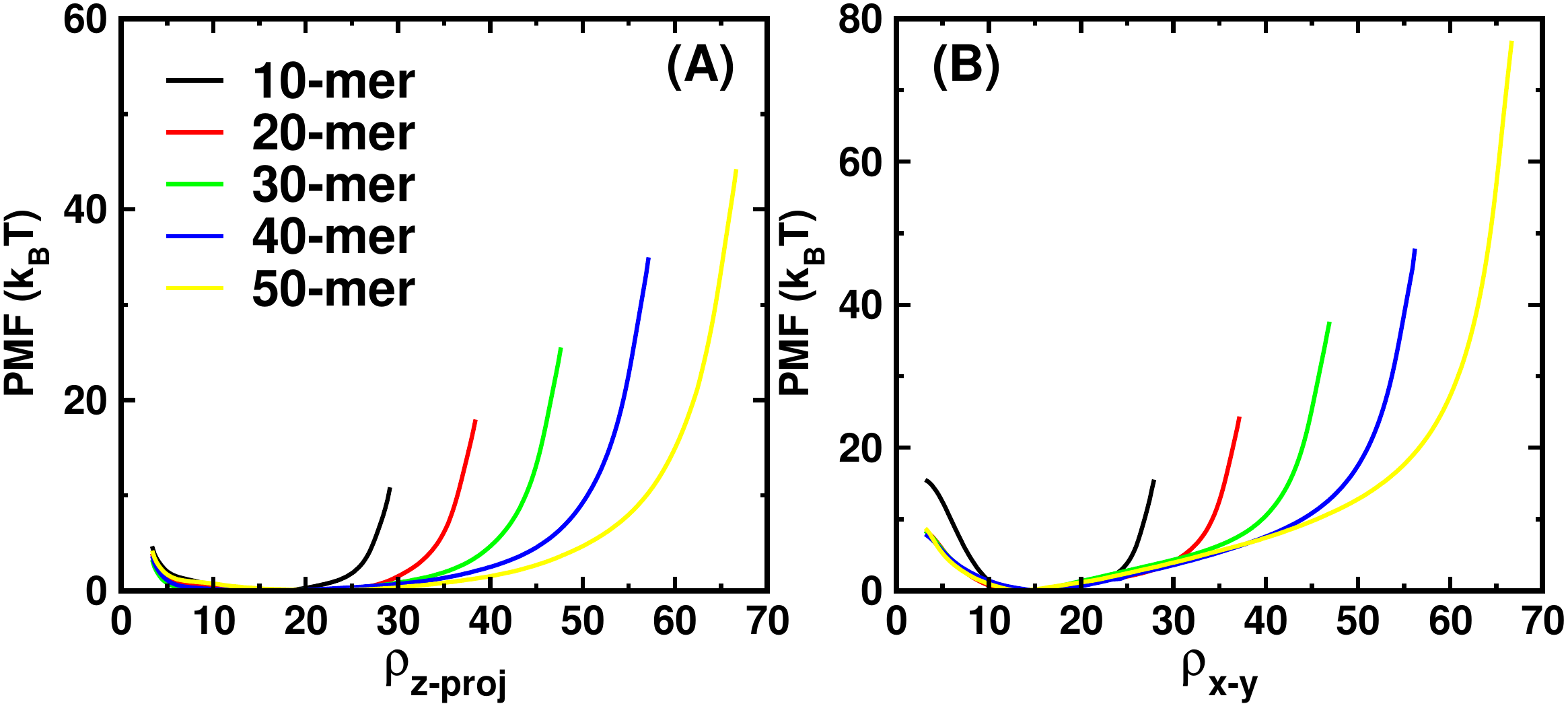}
\includegraphics[width=\columnwidth]{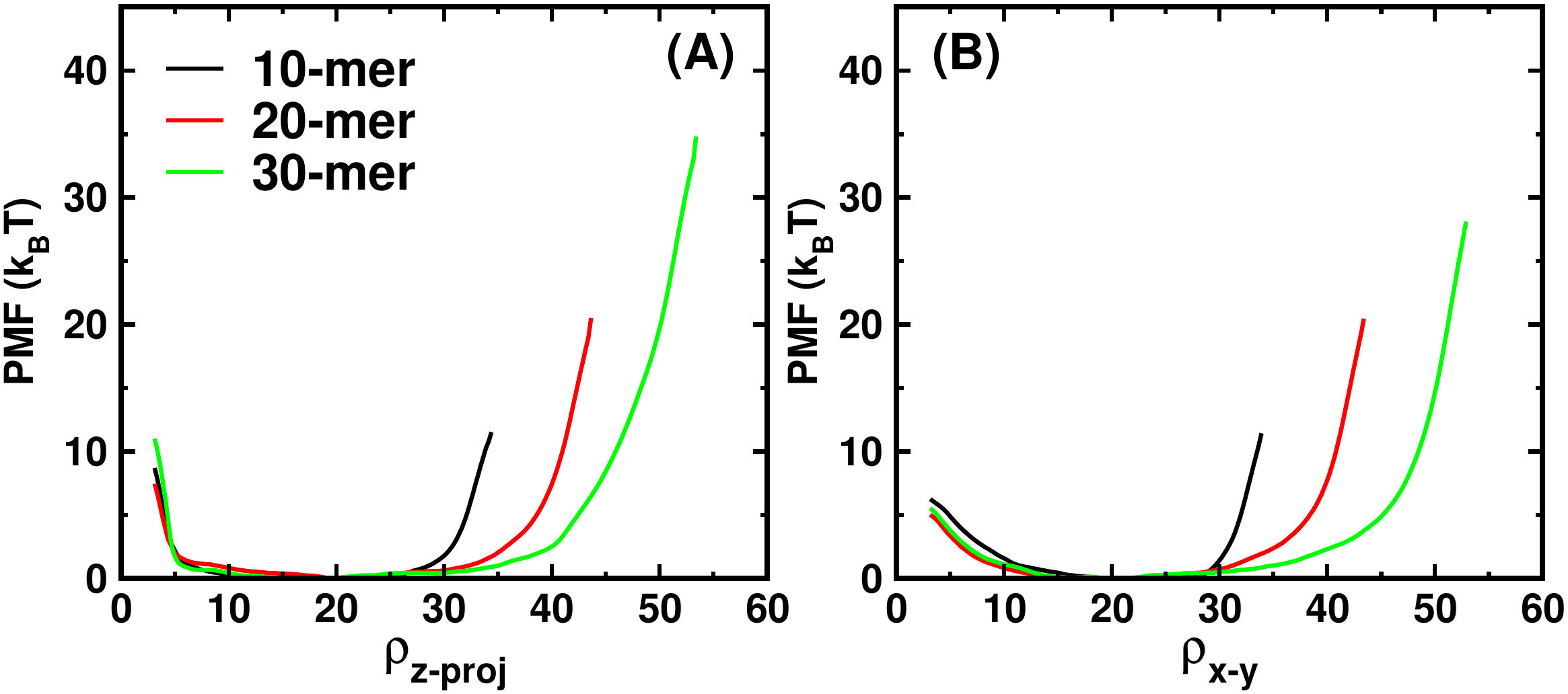}
\caption{(Color online)  \textbf{PMF profiles for tethered diabodies}. PMF profiles as a function of the $z$-projection of 
the vector joining the tethering point to the center 
of mass of the free nanobody for the (A) SPH and (C) SBCG systems. 
PMF profiles as a function of the projection on the $xy$ plane of the vector joining the tethering point to the 
center of mass of the free nanobody, with the free NB restrained to stay at a $z$-projection of 
5$\,\sigma$ for the (B) SPH and (D) SBCG systems.
The $x$-axis is represented in  units of $\sigma$.}
\label{FIG.pmf}
\end{figure}

\noindent The PMF profiles as a function of $\rho_{z-proj}$ are 
shown in Fig.~\ref{FIG.pmf} (A) and (C) for the SPH and SBCG diabodies, respectively. 
In spite of the PMF profiles having similar shape, 
there are a few notable differences. For the 10-mer system, for example, the PMF profile is flat 
in 15$\sigma$ $\le\rho_{z-proj}\le$ 25$\sigma$ for the SBCG system, while the same region for the SPH system occurs 
at shorter distances,  extending between 8$\sigma$ and 20$\sigma$. In addition, for small values of RC, 
the profiles for the SBCG system have a larger slope as compared to the SPH system, 
which means that the SBCG nanobody faces a larger (entropic) repulsive force from the fixed wall than the SPH nanobody. 
The derivative of the PMF profiles is shown in the SI~\cite{suppl}. These differences suggest that 
the different coarse-graining schemes highlight differences in the dynamics near the tethering wall.
In particular, an excessively simple spherical model seems to fall short of capturing important 
features of the interactions with a boundary.    \\
%
%
\begin{figure}
\centering
\includegraphics[width=\columnwidth]{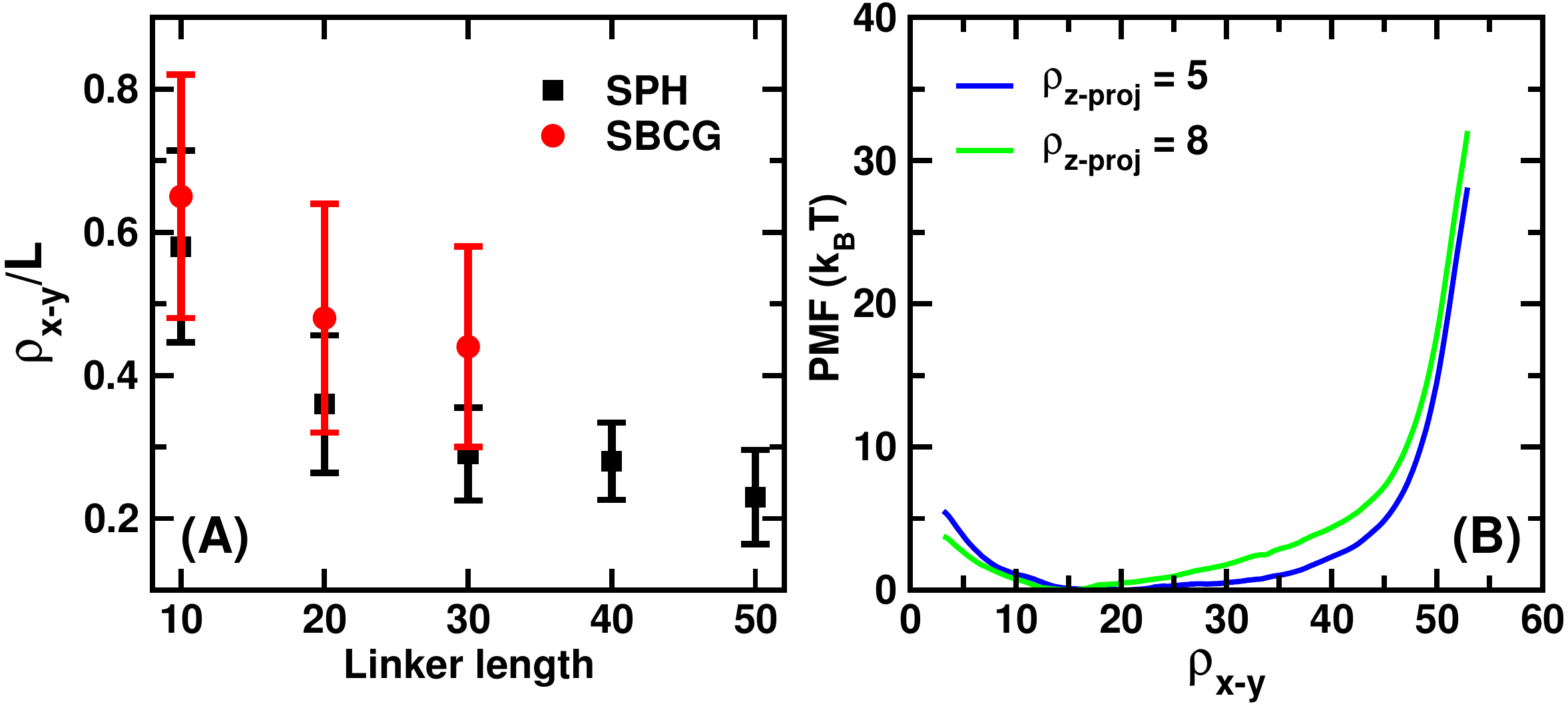}
\caption{(Color online)  (A) Average of $\rho_{x-y}/L$ over the flat region of the PMF profiles as a function of linker length. 
The error bars quantify the extent of the flat region. (B) PMF profiles for the SBCG diabodies with the 30-mer 
linker as a function of $\rho_{x-y}$ for two different values of $\rho_{z-proj}$. 
The PMF profile for a wall-touching situation is different from the one where nbd-1 is farther from the fixed wall. 
In that case, the PMF profile for the SBCG diabody resembles that of the SPH diabody.}  
\label{FIG.comp}
\end{figure}
%
\indent The PMFs are portrayed as a function of $\rho_{x-y}$ in Figs.~\ref{FIG.pmf}~(B) and (D). 
The profiles for the two models appear very different. While the PMFs for the SPH systems increase steadily 
in a linear fashion after $\rho_{x-y}$ $\sim 15\,\sigma$, the profiles for the SBCG systems are flatter and start 
rising at a much later stage. 
Fig.~\ref{FIG.comp}~(A) shows the average of all values of the (normalized) $\rho_{x-y}$ values 
for which PMF$(\rho_{x-y}) \le k_BT$ 
for $\rho_{z-proj}=5 \,\sigma$ as a function of the linker length. The error bars gauge the 
flatness of the PMF profiles, which is seen to vary markedly in the two models.
More precisely, the PMF profiles are flat (in the $xy$ plane) for short linkers, while
the flatness decreases with increasing linker length. This indicates 
that a network of nanobodies linked with short linkers may be more effective at inter-epitope 
binding even for configurations with a large standard deviation in the inter-epitope distances.\\
\indent To better understand the differences displayed by the two models, 
we computed the PMF along $\rho_{x-y}$ for 
different values of $\rho_{z-proj}$ for the SBCG diabodies with 30-mer linkers. 
Fig.~\ref{FIG.comp}~(B) shows the profiles obtained for $\rho_{z-proj}=5 \,\sigma$ 
(nbd-1 in contact with the fixed wall) and $\rho_{z-proj}=8 \,\sigma$ (nbd-1 separated from the wall).
For the larger value of $\rho_{z-proj}$, the PMF profile appears similar to that of the SPH 
diabody, which indicates that the presence of the wall shapes the PMF profiles of the non-spherical model.
We infer that for epitope systems with targets present very close to the cell-membrane, which acts as a soft wall, 
one might have a larger chance of inter-epitope binding as the wall seems to have a flattening effect on the SBCG 
PMF profiles. Overall, the differences between the PMF profiles of the two systems demonstrates 
that molecular shape may play 
a big role in controlling the interaction with the tethering wall.\\
\indent One notices a steeper increase of the PMF as a function of $\rho_{x-y}$ at 
low distances for the 10-mer as compared to other linker lengths.
This effect is much more pronounced for the spherical molecules (Figs.~\ref{FIG.pmf} (B) and (D)). 
This indicates that when the linker length is comparable to the dimensions of the nanobodies, 
it is difficult to approach the epitopes close-by to the one to which the diabody is tethered. 
It is to be stressed that the slope is larger in case of the SPH system as compared to 
the SBCG system, which indicates that the shape of the nanobody is expected to play a role when it comes to 
inter-epitope binding for a high target density,  especially for low linker lengths.\\
\indent In the spirit of computationally aided molecular design, the PMF profiles as a function of $\rho_{x-y}$ 
can be used to estimate the length of the linker required to efficiently result in multi-epitope binding for 
a given target geometry. 
With a knowledge of the epitope size and average inter-epitope distances, one can estimate, with a knowledge 
of the position of minima of the PMF profiles and also their degree of flatness,  what length (or range of lengths) 
of the linker polymer would result in avidity. Simulations with bending modulus of the linker matching different 
polymers used as linkers in practical situations, like various peptides or nucleic acids, 
can  help predict the appropriate stiffness of the linker for a given geometrical arrangement of epitopes. 
Tumor receptors like HER2~\cite{her2} have one epitope for natural antibodies while some engineered triple 
helix proteins called affibodies~\cite{aff1, aff2} are known to engage a different region on the opposite 
side~\cite{her2-aff}. With PMF calculations and knowledge of the position of minima as a function of a
suitable reaction coordinates, one can predict, using MD simulation, the linker length that would maximize 
bispecific binding.
%
\begin{table*}[!bht]
\begin{center}
\caption{\textbf{Simulation details.} Four sets of simulations were performed for calculating the flight/residence times, 
corresponding to different choices of the parameters describing the interaction between CG beads forming the SBCG NBs and
the wall.}\label{tab1}
\begin{tabular}{lllll}\hline
Simulation         & System                                & wall parameter            & wall parameter        & $\qquad
																										   \epsilon$\\
type               &                                       & (SPH nbd1/2)              & \textbf(other beads)  &             \\ 
\hline\hline
%
\textbf{Repulsive} & SPH (10-, 20-, 30-, 40-, 50-, 60-mer) & $s_{wall}=4.5/3.0$        &  $s_{wall}=0.8$       & $\qquad 0$  \\
\textbf{wall \#1}  &                                       & $r_{cutwall} = 5.05/3.36$ &  $r_{cutwall} = 0.89$ &             \\
                   & SBCG (10-, 20-, 30-mer)               &  n. a.                    &  $s_{wall}=0.8$       & $\qquad 0$  \\
                   &                                       &  n. a                     &  $r_{cutwall} = 0.89$ &             \\
\textbf{Repulsive} & SPH (10-, 20-, 30-, 40-, 50-, 60-mer) & $s_{wall}=4.5/4.5$        &  $s_{wall}=0.8$       & $\qquad 0$  \\
\textbf{wall \#2}  &                                       & $r_{cutwall} = 5.05/5.05$ &  $r_{cutwall} = 0.89$ &             \\
                   \\
\hline\hline
\\
\textbf{Attractive} & SPH (10-, 20-, 30-, 40-, 50-, 60-mer) &  $s_{wall}=4.5/4.5$  & $s_{wall}=0.8$  &  $\qquad1.5 \,k_{B}T$\\
\textbf{wall \#1}&& $r_{cutwall}=11.25/11.25$  &  $r_{cutwall} = 2.0$  &   \\
&& \\
\textbf{Attractive} & SPH (10-, 20-, 30-, 40-, 50-, 60-mer) &  $s_{wall}=4.5/4.5$  & $s_{wall}=0.8$  &  $\qquad2.5 \,k_{B}T$\\
\textbf{wall \#2}&& $r_{cutwall}=11.25/11.25$  &  $r_{cutwall} = 2.0$  &    \\
\hline
\end{tabular}
\end{center}
\end{table*}

%
\subsection{$z$-distribution of the free paratope: comparison with models}
%

%
\begin{figure}
\centering
\includegraphics[width=\columnwidth]{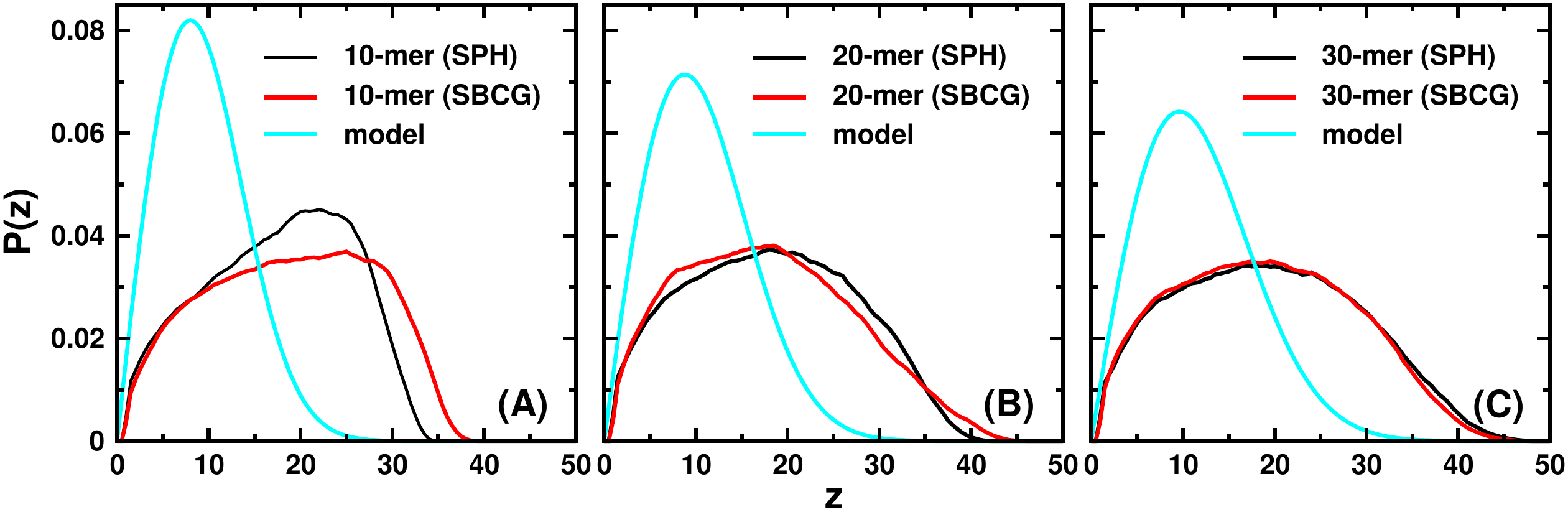}
\caption{(Color online)  $z$-distribution of the free paratope (bead P1) for (A) 10- (B) 20- and (C) 30-mer SPH and SBCG systems 
plotted along with the equivalent Gaussian polymer approximation~\eqref{eqm} (see text).}  
\label{FIG.z-dist-sbcg}
\end{figure}
%

\noindent An important observable that is tightly related to the statistics of flight and residence times 
is the distribution of the $z$-coordinate of the paratope of the 
free nanobody (bead P1). We computed these distributions  
for all the systems from the simulations performed to calculate the flight and residence times (see methods).
It is interesting to compare the data for the SPH and SBCG systems 
with a simple model. The expression for the normalized equilibrium $z$-distribution of the 
free end of a Gaussian polymer, tethered at a height $z_0$ from a reflecting wall, 
reads~\cite{book}
\begin{widetext}
\begin{equation}
P(z) = \sqrt{\frac{3}{2\pi N_kb^2}}\frac{1}{{\rm erf} \left(z_0\sqrt{3/2N_kb^2}  \right)}
       \left[ 
            e^{-3(z-z_0)^2/2N_kb^2} - e^{-3(z+z_0)^2/2N_kb^2}
       \right]
\label{eqm}
\end{equation}
\end{widetext}
where $N_k$ is the number of Kuhn monomers, $b$ is the Kuhn length and $z_0$ is the $z$-coordinate of the 
tethering point measured from the wall.
In the limit $z_0\to$ 0, the expression reduces to~\cite{marzio}:
\begin{equation}
P(z) = \frac{3z}{\langle d^2\rangle}  \, e^{-3z^2 / 2\langle d^2\rangle }
\label{eqn}
\end{equation}
where we have introduced explicitly the  
average square end-to-end distance of the free polymer, 
$\langle d^2\rangle = N_kb^2$.\\
\indent It is interesting to inquire whether the data for the composite two-NB models can be 
described by an {\em effective} polymer model. 
The most obvious choice would be a Gaussian polymer with the same flexibility as the diabody linker,
contour length equal to the contour length of the diabody measured from CB2 to the free 
paratope (P1) and tethered at the average height of CB2 above the lower $z$ wall. 
A reasonable guess for the length of the effective model is thus $N_k$ = $N+2r/b$, 
where $N$ is the number of monomers in the linker, $r$ is half 
the distance between P1 and CB1 and $b$ is the Kuhn length of the linker polymer 
(see again Fig.~\ref{FIG.bead-name}). 
We set $z_0 = 7.5 \,\sigma$,  which is the average height for CB2 
measured from the wall for the SPH system (see SI~\cite{suppl}). 
This would represent a polymer tethered at 
$z_0 = 7.5 \, \sigma$ distance units above the tethering wall and 
having a contour length equal to the linker and nbd1 combined. 
%
%
Our linker has a persistence length of $\ell_p \simeq 1.5 \, \sigma$ (see SI),
which leads to a Kuhn length $b = 2\ell_p = 3 \, \sigma$. \\
\indent Fig.\ref{FIG.z-dist-sbcg} shows the comparison of the effective model with the data 
for the SPH and the SBCG systems.
It is apparent that the distribution for the diabodies describes a higher representation of 
large $z$ values and a  reduced representation of small $z$  values when compared to the tethered polymer. 
This is an expected consequence of the higher entropic repulsive force exerted by the wall on  
the free bead, a mechanism akin to the entropic pulling force demonstrated in the disassembling 
and translocating action of Hsp70s chaperones~\cite{entropicpull}.
More precisely, the conformational space near the wall is restricted more severely for the diabodies, due to
presence of the nanobodies at the two ends of the polymer, than for a {\em bare} polymer with the same contour 
length and flexibility.
The extent of the difference between the simulation and the model suggests extended, composite molecules such 
as our diabodies, belong to a different universality class altogether. One can, however, 
use eq.~\eqref{eqm} or~\eqref{eqn} 
to determine the effective Kuhn length $N_k^{\rm eff}$ of an effective polymer with the same persistence length $\ell_p$ 
as the linker in the diabodies.  For this, we fit the distributions of the 40-, 50- and 60-mer SPH 
systems with eq.~\eqref{eqn} (see SI) with $b = 3\,\sigma$, which leads 
to $N_k^{\rm eff} \sim 108$, $115$ and $119$ for 40-, 50- and 60-mer diabodies respectively.\\
\indent It is interesting to note that the distributions for the SBCG and the SPH systems differ to a highest extent 
when the linker length matches the dimensions of the linked NBs, i.e. for 
the shortest linker (10-mer), while they approach each other rapidly as 
the linker length increases and almost overlap for the 30-mer linker. This means that, as the statistics of the
vertical coordinate above the wall is concerned, for linkers longer than approximately the size of the NBs, an 
equivalent SPH model can be used, entailing considerable simplification of the simulations.

\subsection{Flight/residence times: quantifying the kinetics of the second binding}

\noindent The statistics of the flight/residence times (F/RT) 
of the free NB are crucial observables, as they embody the kinetic determinants of the second binding, 
hence can help quantify avidity.
The F/RTs are defined as stretches of consecutive frames that the  paratope of the free NB (bead P1) 
spends above/below, respectively, a fixed threshold height $z_{th}$ from the tethering wall. 
Let us denote with $P_f(t)$ and $P_r(t)$ the equilibrium distributions of flight and residence  times.
The corresponding 
complementary cumulative distributions, $S_f(t)$ and $S_r(t)$, defined as 
\begin{eqnarray}
\label{e:Sfr}
&&S_f(t) = \int_t^\infty P_f(t^\prime)\,dt^\prime \nonumber\\
&&S_r(t) = \int_t^\infty P_r(t^\prime)\,dt^\prime 
\end{eqnarray}
coincide with the survival probabilities relating to the corresponding domains,
$\mathcal{D}_f = \{z \in [0,L] \, | \, z \ge z_{th} \}$ and  
$\mathcal{D}_r = \{z \in [0,L] \, | \, z < z_{th} \}$,
$L$ being the side of the simulation box along the $z$ direction.
We note that the domain  $\mathcal{D}_r$ can also be regarded as the paratope-wall {\em interaction} domain. 
More precisely, $S_f(t)$ represents the fraction of flight events whose duration exceeds $t$, hence this is 
nothing but the probability that  the paratope be still in region $\mathcal{D}_f$ after a time $t$,
i.e. indeed the survival probability for domain  $\mathcal{D}_f$. Analogously,  $S_r(t)$ represents
the survival probability for domain  $\mathcal{D}_r$. The corresponding distribution of exit times 
(in the sense of first passage times), $\mathcal{P}_f(t)$ and $\mathcal{P}_r(t)$, 
can then be computed straightforwardly as~\cite{MFPT}
\begin{eqnarray}
\label{e:Pfptfr}
&&\mathcal{P}_f(t) = -\frac{dS_f(t)}{dt} \nonumber\\
&&\mathcal{P}_r(t) = -\frac{dS_r(t)}{dt} 
\end{eqnarray}
The inverse mean-exit times can be considered as measures of the escape rate from the corresponding domains.
Therefore, combining eqs.~\eqref{e:Sfr} and~\eqref{e:Pfptfr}, it is possible to estimate the on
and off rates directly from the series of flight and residence times, as
\begin{eqnarray}
&&k_{\rm on}  = \left[ \int_0^\infty t \,\mathcal{P}_f(t) \, dt\right]^{-1}  
              = \left[ \int_0^\infty S_f(t) \, dt              \right]^{-1}  \label{e:kon}\\
&&k_{\rm off} = \left[ \int_0^\infty t \,\mathcal{P}_r(t) \, dt\right]^{-1}  
              = \left[ \int_0^\infty S_r(t) \, dt              \right]^{-1}  \label{e:koff}
\end{eqnarray}
To calculate the flight ($t_f$) and residence times ($t_r$) numerically, the $z$-coordinate of the paratope was 
monitored  through the simulation. A threshold $z$-height of
$z_{th} = 3.5\,\sigma$ was set to define whether the paratope is in the 
flight or residence regions. If the paratope stayed above or below the threshold for at least 0.5 ns, 
the corresponding trajectory stretch was designated to correspond to a flight or residence event, 
respectively. This additional constraint was tailored specifically to avoid short recrossing events that 
could bias the statistics unphysically at short times. 
Data for such events were accumulated over 30 $\mu$s-long trajectories of 25 non-interacting tethered diabodies. 
A set of 3 such simulations were performed for each linker length. In addition, the simulations were performed for 
both repulsive and attractive tethering walls (with attractive energies equal to 1.5~$k_BT$ and 2.5~$k_BT$),
with the aim of assessing the effect on the paratope-wall kinetics 
of some weak non-specific attraction between the protein and the wall/membrane.\\
%
%
\begin{figure}[t!]
\centering
\includegraphics[width=\columnwidth]{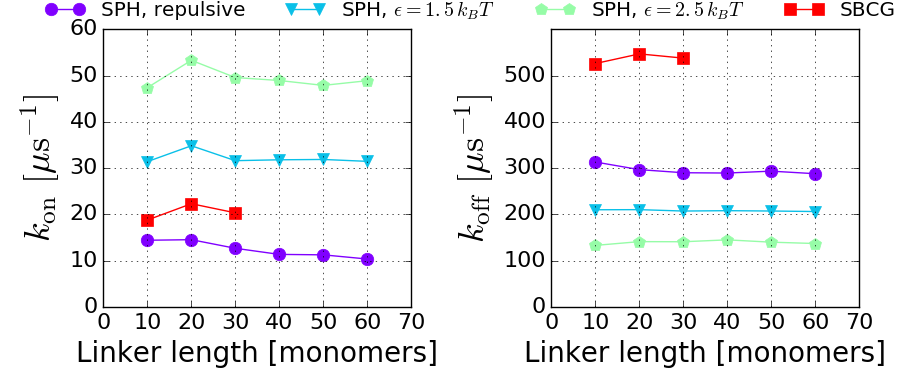}
\includegraphics[width=\columnwidth]{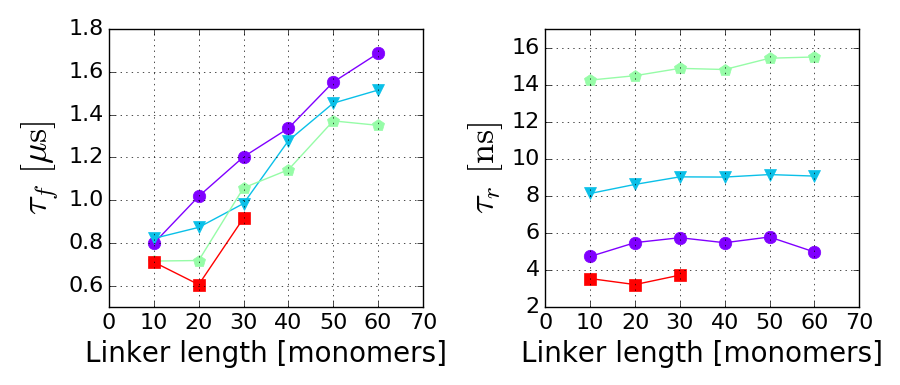}
\caption{(Color online) Upper panels. 
On and off rates describing the kinetics of paratope entering and exiting the interaction 
domain $\mathcal{D}_r$ (i.e. $z<z_{th}$), computed according to the prescriptions~\eqref{e:kon} and~\eqref{e:koff}.
Bottom panels. Time constants of the exponential tails of the survival probabilities. These 
can be thought of as the average time of the rare, very long events.
The data for the SPH model refer to the simulation set \#2 (see Table I)}  
\label{f:konoff}
\end{figure}
%
%
%
\begin{figure*}[t!]
\centering
\includegraphics[width=1.5\columnwidth]{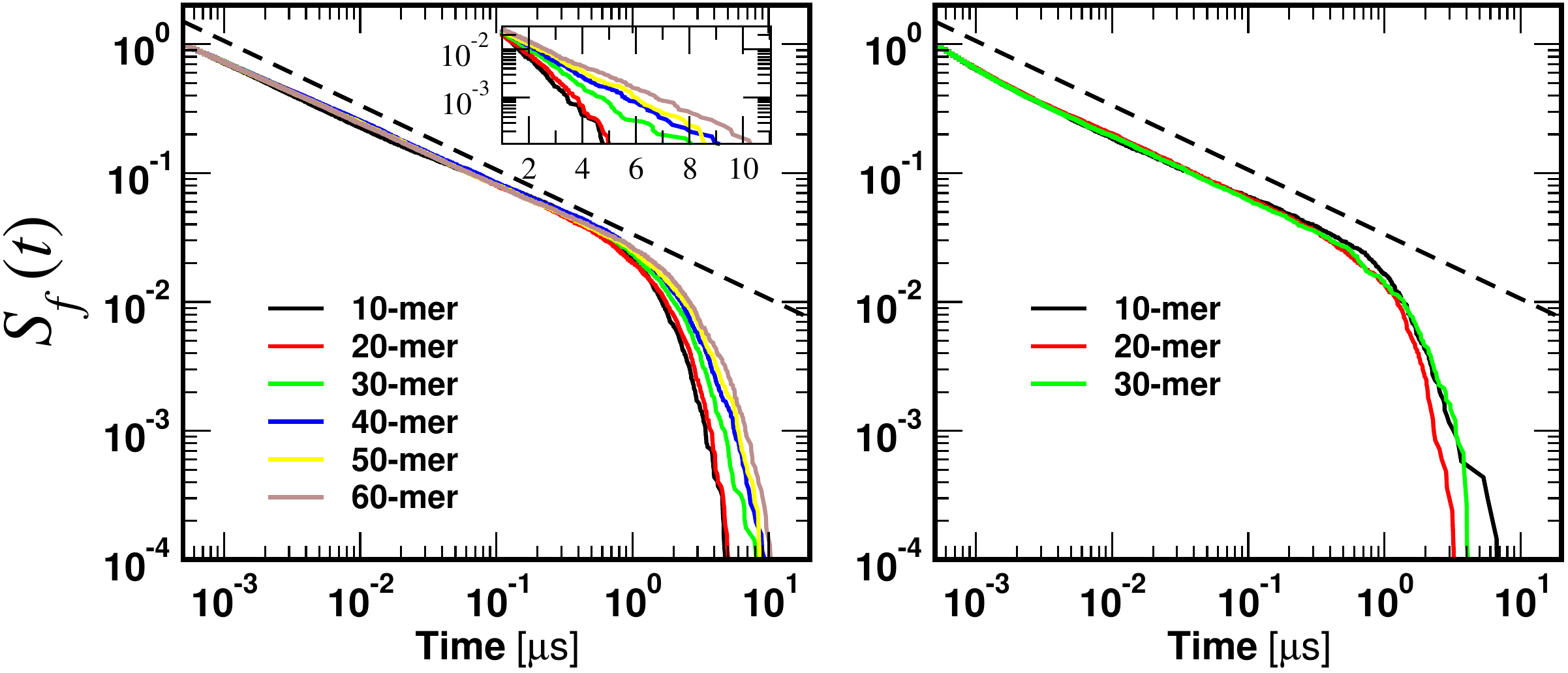}
\includegraphics[width=1.5\columnwidth]{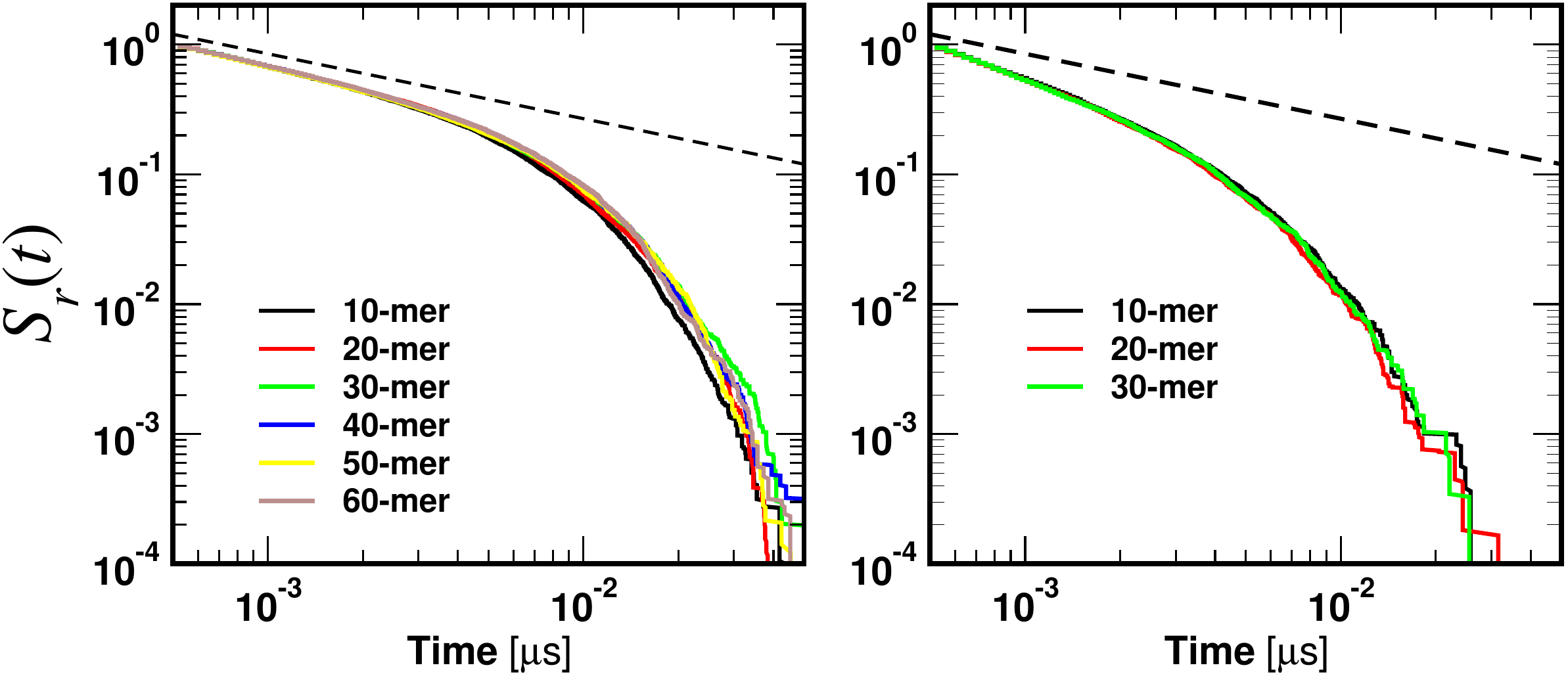}
\caption{(Color online)  Survival probability of the epitope in the flight ($z\ge z_{th}$, top)
and residence ($z< z_{th}$, bottom) domains for 
(left) the SPH system and (right) SBCG system for different linker lengths. 
The dashed lines are plots of power laws of the kind $t^{-1/2}$.
The inset shows a close-up in lin-log scale of the tails of the flight 
survival probability, which makes the exponential decay clearly visible.
The plots for the attractive walls are provided in the SI~\cite{suppl}.}  
\label{FIG.frac}
\end{figure*}
%
\indent Fig.~\ref{f:konoff} (upper panels) illustrates the calculation of the on and off rates 
as described by formulas~\eqref{e:kon} and~\eqref{e:koff}. The probability per unit time that the paratope
enter the interaction domain appears of the order of  tens of $\mu$s$^{-1}$, while the 
probability per unit time that it exit the same domain turns out to be about ten times higher.
Interestingly, the SBCG model shows a higher on-rate than the spherical model (with a pure repulsive 
wall). At the same time, the exit probability is higher for the SBCG model. \\
\indent Fig.~\ref{f:konoff} also demonstrates that a weak non-specific attraction 
between the NBs and the wall of the order $\epsilon \simeq 1-2 \, k_BT$
increases the on-rate (i.e. decreases the overall survival probability in the flight domain) 
and decreases the off-rate (i.e. makes journeys of the paratope in the interaction domains longer). 
More specifically, these data refer to a modified SPH-wall system with non-specific isotropic 
attraction (LJ) between the wall and nbd-1/2 and P1/2. 
It is interesting to observe that the gain afforded by a weak attractive wall in terms 
of on-rate, as gauged by $k_{\rm on}(\epsilon,N)/k_{\rm on}(0,N)$, is found to increase with the linker 
length $N$. In fact, while, $k_{\rm on}(0,N)$ decreases with $N$, a weak attraction 
makes $k_{\rm on}(\epsilon,N)$ nearly insensitive to variations in the linker length. 
This possibly reflects the fact that the entropic cost associated with  entering the interaction 
domain decreases with increasing linker length in the presence of attraction between the paratope/NB system
and the wall. Conversely, the ratio $k_{\rm off}(\epsilon,N)/k_{\rm off}(0,N)$ appears to remain 
constant as $N$ increases.\\
\indent It is instructive to inspect in more detail the survival probabilities. 
Fig.~\ref{FIG.frac} depicts the survival curves for both the flight and paratope-wall 
interaction  domain. The short time behavior ($t \lesssim 0.5$ $\mu$s) appears to follow an inverse power 
law with exponent $1/2$ (see dashed lines in Fig.~\ref{FIG.frac}), irrespective of the linker length. 
This is the expected behaviour for the survival probability of a free random walk in three dimensions~\cite{MFPT}. 
This means that short survival times in either domain are dominated by the unconstrained diffusion 
of the paratope. By contrast, longer survival events depend markedly on the length of the linker
and are distributed exponentially. The inset in Fig.~\ref{FIG.frac} makes this point very clear in 
the case of the function $S_f(t)$ for the SPH model. We find that this is a general feature of the 
tails of the paratope survival probabilities in either region. As it shows from the figures, 
the tails of the flight times depend on the linker length, while those of the residence times 
much less so. In order to have some insight into the tail, rare-event region, we might 
reason as follows. \\
\indent We can safely assume that rare events make uncorrelated time series. In this case, the 
statistics of return events will be specified by the configurational probability (independent of time) that the paratope 
be in the relevant regions, either $z \geq z_{th}$ or $z < z_{th}$. In turn, this will depend on the conformational 
statistics of the NB-polymer systems and, in the absence of an appropriate analytical model, 
can be easily determined from our PMF calculations (see Fig.~\ref{FIG.pmf}).
Let us denote with $P_>$, $P_<$ the equilibrium probability that the free NB be above or below the threshold $z_{th}$,
respectively. In this case, the probability $P_a(k)$ and $P_b(k)$ 
of observing $k$ consecutive sampled frames above (a) or below (b) $z_{th}$, respectively, can be computed as 
\begin{eqnarray}
&&P_a(k) = P_> (1-P_<)^{k-1}P_< \simeq  P_{<}\,e^{-P_<k} \label{e:Pak}\\
&&P_b(k) = P_< (1-P_>)^{k-1}P_> \simeq  P_{>}\,e^{-P_>k} \label{e:Pbk}
\end{eqnarray} 
If for the sake of the argument we take the Gaussian tethered model~\eqref{eqn} as a reference case, 
it is readily seen that
\begin{eqnarray}
&&P_{>} = \int_{z_0}^{+\infty}P(z) \, dz = e^{-3z_0^2/2\langle d^2\rangle}   \label{e:Pgt}\\
&&P_{<} = \int_{0}^{z_0}P(z) \, dz       = 1-e^{-3z_0^2/2\langle d^2\rangle}       \label{e:Plt}
\end{eqnarray}
If we introduce the time decay constants $\tau_f$, $\tau_r$ of the exponential tails 
of the survival probabilities in the flight and interaction domains, respectively,
Eqs.~\eqref{e:Pgt} and~\eqref{e:Plt} entail
\begin{eqnarray}
&&\tau_f = \frac{\Delta t_c}{P_<} = \frac{\Delta t_c}{1 - e^{-3z_0^2/2\langle d^2\rangle}}
                       \simeq  \Delta t_c \frac{2\langle d^2\rangle}{3z_0^2}  \label{e:tauf} \\
&&\tau_r = \frac{\Delta t_c}{P_>} = \Delta t_c  \, e^{3z_0^2/2\langle d^2\rangle} 
                       \simeq  \Delta t_c \left( 1 + \frac{3z_0^2}{2\langle d^2\rangle} \right)
                       \label{e:taur}
\end{eqnarray}
where $\Delta t_c$ is  a time of the order of the typical 
correlation time of consecutive frames  and in the last passages we have made use 
of the fact that $z_0^2/\langle d^2\rangle \ll 1$.\\
\indent Fig.~\ref{f:konoff} indeed shows that $\tau_f$ increases linearly 
with $N$ ($\langle d^2\rangle\propto N$) for the spherical 
model, according to the prediction~\eqref{e:tauf}. It is interesting to observe that the slope does not
seem to depend on the value of the weak attractive energy $\epsilon$. This is expected, as
rare, long flight events are dominated by the statistical weight of the
configurations of the combined NB/linker molecule away from the wall. 
The simple calculation leading to eq.~\eqref{e:taur} also correctly
explains the observed reduced variability of residence times as the linker 
length is increased (see again Fig.~\ref{f:konoff}, bottom right panel). 
However, a closer inspection reveals that the average duration of rare, long residence events 
increases with  the linker length $N$ in the presence of an attractive interaction between
the paratope/NB system and the wall, even though with a much smaller slope than for the increase of $\tau_f$. 
Overall, we conclude that the effect of a weak non-specific interaction with the wall
decreases the average duration of rare, long flights and residence events. 
It is interesting to observe that the SBCG model (with a repulsive wall)  
displays the shortest duration of rare long flights (Fig.~\ref{f:konoff}, bottom left), 
even shorter than for the spherical model in the presence of the most attractive wall ($\epsilon = 2.5 k_BT$). 
Moreover, there seems to be an optimum (a pronounced dip) at a linker length that is approximately 
the same size of the attached NBs ($N=20$).\\
\indent While rare, long flight and residence times are on the $\mu$s and 
tens of ns scales (exponential tails), 
respectively, the average values $\langle t_f \rangle$, $\langle t_r \rangle$ 
are dominated by the short-time power-law behaviour.
Fig.~\ref{FIG.flt} reveals that average flight times turn out to be of the order of 
about $10^2$ ns, while residence times are about 50 times shorter, of the order of $2-3$ ns. 
Furthermore, one can appreciate that the SBCG model systematically displays
shorter flight and residence times with respect to the spherical model. As for rare events, this feature should be 
attributed to the shape and intrinsic flexibility of the SBCG NBs as compared to equivalent 
rigid spheres of the same size. \\
\indent It is interesting to observe that shorter linker (10-mers) correspond to rather unfavorable 
situations (large flight times). Remarkably, increasing the linker length from $N=10$ results in a 
substantial reduction in the duration of the flight stretches, 
an effect that is more pronounced for the SBCG model (see again Fig.~\ref{FIG.flt} A). 
A close inspection of the trajectories shows that for the 10-mer linker, for which the linker length 
is less than the size of the nanobodies, the tethered unit, due to its steric extension, 
introduces a steep entropic barrier for the free NB when it approaches the wall (see SI~\cite{suppl}).
The consequence of such steric repulsion and its more pronounced effects in the case of the 10-mer linkers 
can also be seen from Fig.~\ref{FIG.z-dist-sbcg}, where the peak of the $z$-distributions of the free paratope (bead P1) 
are  farthest away from that of the simple equivalent Gaussian polymer for the 10-mer, 
and more so for the SBCG diabody.  This effect is present for the SPH model too,
albeit less marked. Again, also for average values, there seems to be an {\em optimum} length of 
the linker that minimizes the average time spent by the free NBs away from the tethering surface, that 
approximately matches the size of the binding units themselves.
All the data for $t_f$ and $t_r$ for this set of simulations  
are reported in tabulated form in tables I to V in the SI.\\
%
%
\begin{figure}
\centering
\includegraphics[width=\columnwidth]{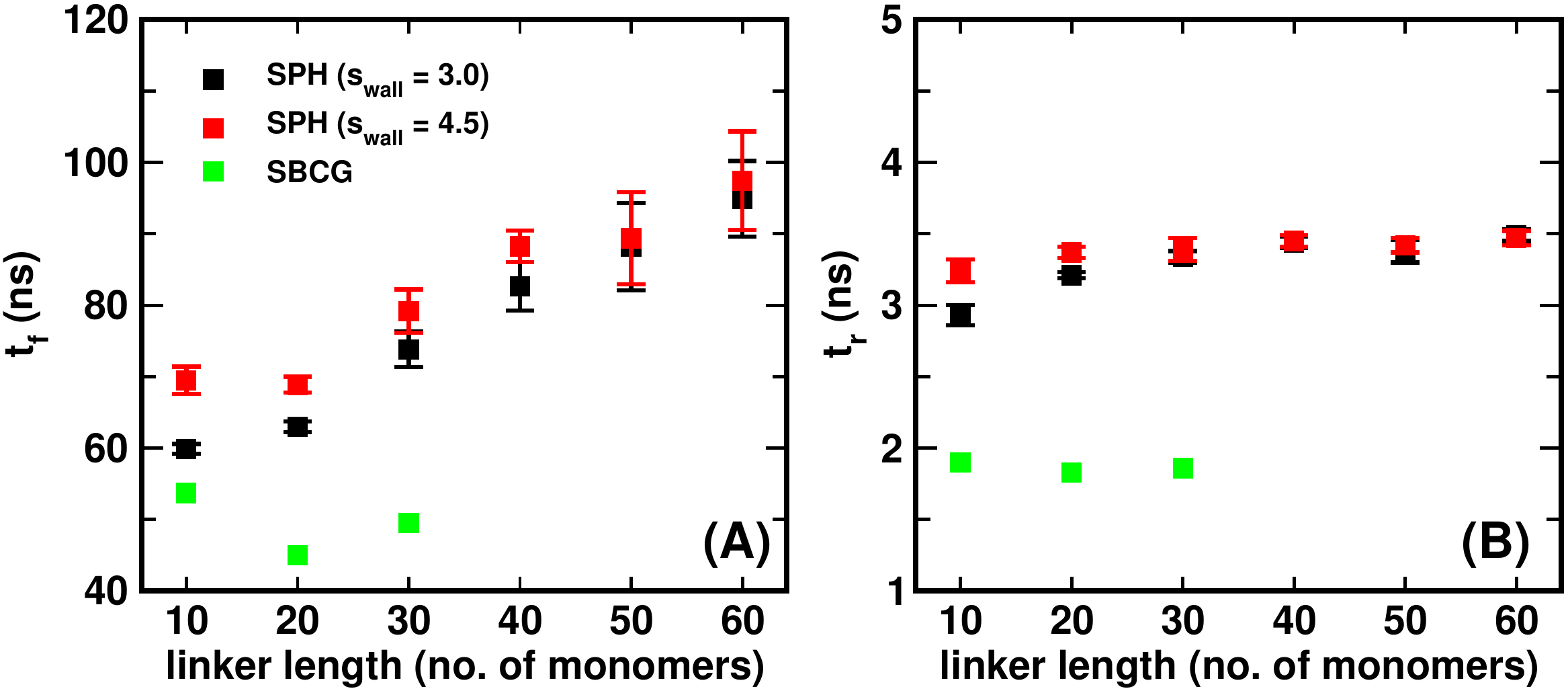}
\caption{(Color online)   Average flight times (A) and average residence times (B), computed 
over the trajectories of 25 non-interacting, tethered diabodies. 
A distance cutoff $z_0 = 3.5\,\sigma$ between the tethering wall and the free paratope (P1) was 
used to calculate the values. A jump above (a plunge below) the cutoff was considered to be a 
flight (residence) event only if it lasted for more than 0.5 ns,  
The systems compared comprise two different variants of the SPH model,
as described in table~\ref{tab1},  and the SBCG system. }  
\label{FIG.flt}
\end{figure}
%
\begin{figure}
\centering
\includegraphics[width=\columnwidth]{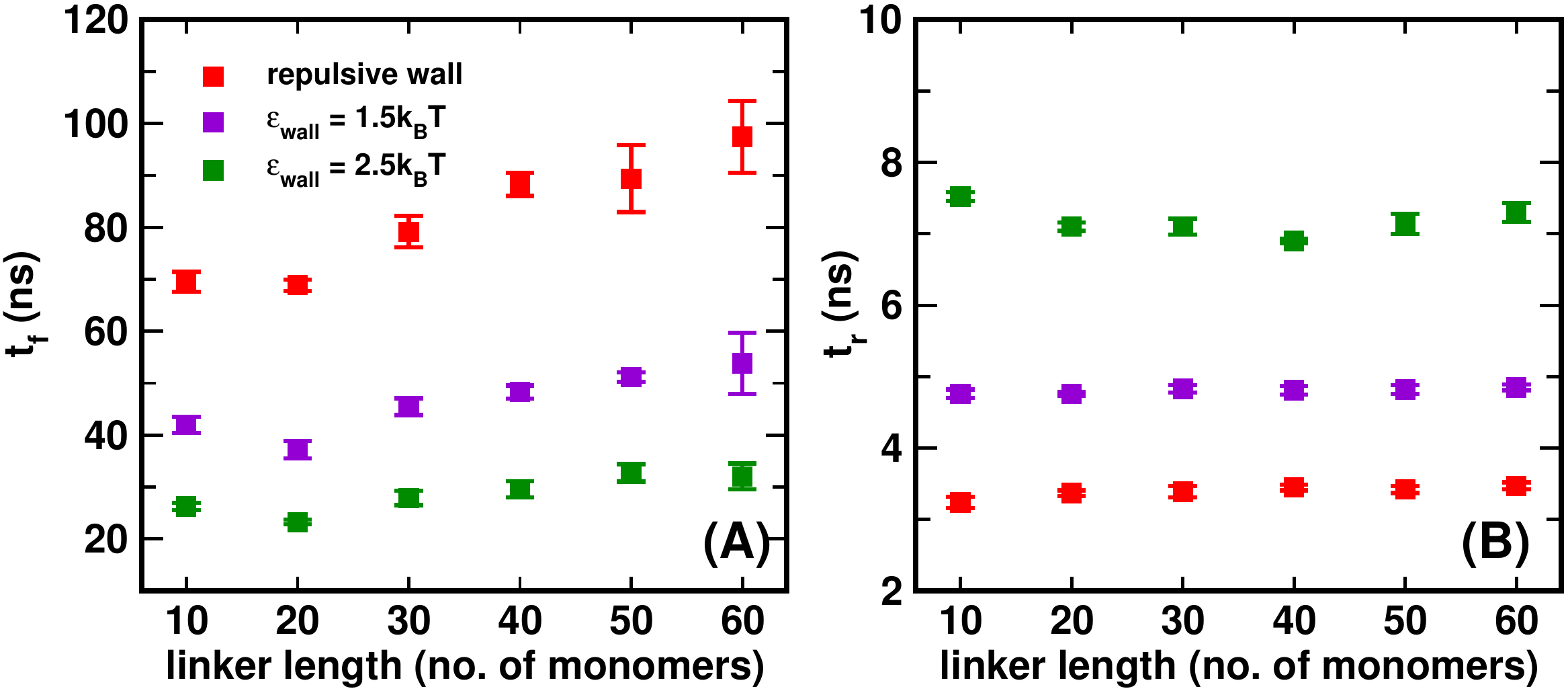}
\caption{(Color online) \textbf{Repulsive vs. attractive wall}. Comparison of (A) 
average flight times and (B) average residence times for SPH systems 
for repulsive and attractive tethering walls computed as described 
in Fig.~\ref{FIG.flt}.}  
\label{FIG.flteps}
\end{figure}
%
%
\indent In order to investigate further the details of the NB-wall interactions, 
we performed the simulations with repulsive tethering walls for two different 
values of the LJ repulsive length $s_{wall}$ ($4.5\,\sigma$ and $3.0\,\sigma$) for nbd-2 (spherical model). 
We observe that for smaller linker lengths (10$-$30-mers), 
flight times are shorter for a smaller value of $s_{wall}$. 
The effect is negligible for the longer linkers, and seems, in fact, inversely proportional to the linker length. 
The effect arises because the value of $s_{wall}$ determines how nbd-2 (the tethered NB) 
would interact with the tethering wall and how much it is able to bend. 
This is expected to affect the value of $\langle t_f \rangle$,
more so in presence of the external harmonic potential that 
restrains the value of the angle between P2-CB2-L1 (the axis of the tethered sphere) 
to 180$^{\rm o}$. One would then expect that how much nbd-2 can bend would depend on the shape of 
the nanobody, as the shape would then determine the dynamics and effective interaction of the 
nanobody with the tethering wall. \\
\indent The bending propensity of the tethered unit can be gauged by calculating the height distribution of 
CB2 (see fig.~\ref{FIG.bead-name}) measured from the tethering wall surface for the SPH  and 
the SBCG systems (see SI~\cite{suppl}). 
From the plot we see that the very low heights have a significant probability in case of the SBCG 
diabodies, which may contribute to the lower values of $\langle t_f \rangle$. 
By contrast, for the SPH diabodies there is 
an abrupt lower cutoff depending on the value of $s_{wall}$.  
From Fig.~\ref{FIG.flt} we notice that the values of $\langle t_f \rangle$ for 
the SBCG diabodies are substantially shorter as compared to the SPH diabodies. For the 30-mer linker, for example, 
the SPH system has  $\langle t_f \rangle \sim 80$ ns while for the SBCG system 
$\langle t_f \rangle  \sim 50$ ns. It thus seems that it is not an obvious task to 
reproduce the interaction with a wall within by a model that preserves the shape of the atomistic structure of the NB
via a simple spherical representation. Thus the SBCG model seems more appropriate to calculate relevant dynamical 
parameters, such as the on-rate for second binding, that are expected to rely substantially on the details 
of the interactions with a wall/membrane. \\
\indent A closer look at the average residence times  reveals that, while  $\langle t_r \rangle$ 
shows negligible dependence on the linker length, the dependence on the model is noticeable. 
The residence time for the SPH diabodies is $\sim 3.5$ ns, while the values for the SBCG diabodies is close to 2 ns. 
The difference can be attributed to the fact that the SBCG nanobody likely generates larger reaction forces from the 
tethering walls on behalf of its fluctuating structure (see SI~\cite{suppl}), 
thus reducing the time it stays near the wall. By contrast, the SPH nanobody would face a smaller reaction 
from the wall, given its rigid and fixed surface. Here again one can appreciate the 
importance of the model being used. \\
\indent Finally, to ascertain that the  length of the flight-residence times simulations 
was sufficient to arrive at converged values of $\langle t_f\rangle$ and $\langle t_r\rangle$, 
we performed the calculation for different durations of the simulations, and checked how the calculated values 
changed as a function of the simulation length. For the 30-mer linker, for example, $\langle t_f\rangle$ started from a 
value of 85 $\pm$ 22 ns for a simulation length of 3.75 $\mu$s and slowly converged to the reported value of $\sim$ 77 ns 
for a simulation length of 15 $\mu$s and stayed close to that for longer simulation lengths, suggesting that our 
simulation length of 30 $\mu$s is appropriate for producing converged results. The values of $\langle t_f\rangle$ 
as a function of simulation length for the SPH systems with $N=30$ and $N = 60$ linkers are shown 
in the SI~\cite{suppl}. 
%

\section{\label{sec:concl}Conclusions and perspectives}

\noindent In this work  we have introduced two different coarse-grained models of polymer-linked, two-nanobody
molecules, as a simple but paradigmatic example of novel immunotherapy agents that are increasingly being developed
in a variety of contexts. More precisely, while the linker has been modeled invariably as a bead-and-spring system
(stretching and bending), the nanobodies have been represened either as 
single large rigid spheres, or as collections of small spheres suitably connected by springs.
Such representation  was parameterized in a bottom-up philosophy directly
from atomistic simulations in explicit solvent. The latter scheme led to binding units displaying 
the same {\em shape} and large-scale flexibility as the atomistic systems.\\
\indent The aim of this work was to lay the bases of coarse-grained, computationally aided
drug design in this area from the firm standpoint of statistical and computational physics. 
In the spirit of the accepted model of sequential binding of bivalent molecules~\cite{iggmio}, 
whereby bivalent agents first bind to a target-covered surface from the bulk, and subsequently 
dynamically explore the surroundings 
for a second target within reach (either on the same or on a facing surface), 
our main focus was to elucidate the physics of the latter kinetic step.
For this purpose, we have mainly focussed on the kinetic and equilibrium properties of a 
molecule tethered to a wall through one of its binding units (NB), in order to investigate 
the main dynamical and structural determinants of the second binding.
In particular, we aimed at investigating (i)  to what extent the degree of coarse-graining 
may impact the dynamics of the combined linker/free NB system and (ii) the interaction dynamics of 
the free paratope (the active binding site carried by the NB) with the surface.  \\
\indent In the first part of this work, we have shown that the calculation of potentials of mean force (PMF)
along specific, one-dimensional collective coordinates can provide considerable insight 
into the nature, strength and range of the thermodynamics forces that govern the motion of the free paratope. 
Furthermore, such calculations 
may constitute a precious tool to investigate the role of such forces in the dynamics of the paratope-wall 
encounter, which, in turn, governs the kinetics of the second binding. We aim at illustrating this aspect 
in a forthcoming publication. For example, the PMF can be fruitfully used as an effective potential 
in approaches based on the Smoluchowski equation or on first-passage processes~\cite{MFPT}.   \\ 
\indent In the second part of the paper, we have delved into the kinetics of the paratope-wall encounters.
To this end, we have developed a general strategy based on dissecting an equilibrium trajectory 
of the free paratope in flight and residence stretches, depending on whether the active site on the free 
NB was found above or below, respectively, 
a critical interaction threshold close to the wall in the vertical direction. 
We have shown that the encounter and escape kinetics with respect to the wall are simply 
related to the survival probability of the paratope in the flight and interaction domains, 
respectively, which can be simply computed from the 
series of flight and residence times observed over a long MD trajectory. 
We have illustrated how this method allows one to estimate the kinetic constants  of the second binding,
$k_{\rm on}$ and  $k_{\rm off}$, in the presence of a purely repulsive wall and with 
weak, non-specific interactions between the free NB and the wall.
Our simple method not only allows one to quantify the role of factors such as the linker 
length and flexibility (not considered in this study) and non-specific interactions on 
the average flight/residence times. It also makes clear and quantify the role and weight  
of rare, long-duration events that show up in the exponential tails of the survival probabilities. \\
\indent It is worth stressing that the statistics of rare events is by 
no means a secondary issue in this context, as in many real-life situations 
the binding kinetics of such molecules is expected to be dominated by fluctuations, e.g. due to low 
copy numbers or tiny reaction volumes. For example, this is the case of novel bivalent and bispecific 
diabodies engineered to bind within the immune synapse, i.e. at the interface of two facing membranes,
on the effector cell (NK or B-cell) on one side and on epitopes on a tumor cell on the other side. 
The synapse covering an area of the order of $100$ $\mu$m$^2$ for a cell-cell separation 
of about 15 nm~\cite{bongrand,nk},
the role of fluctuations in the number of bridging molecules is expected to be important, which likely
makes the statistics of rare events a major determinant of the binding kinetics.

\section{Acknowledgement}
The authors acknowledge the French ministry of higher education under the PIA2 project BIOS for funding.


\newpage

\appendix
\newpage\section{Supplementary Information (SI)}

\subsection{\label{sec:level1}The angle-bending coefficient for the coarse-grained model of the linker}

\begin{figure}[ht!]
\centering
\includegraphics[width=\columnwidth]{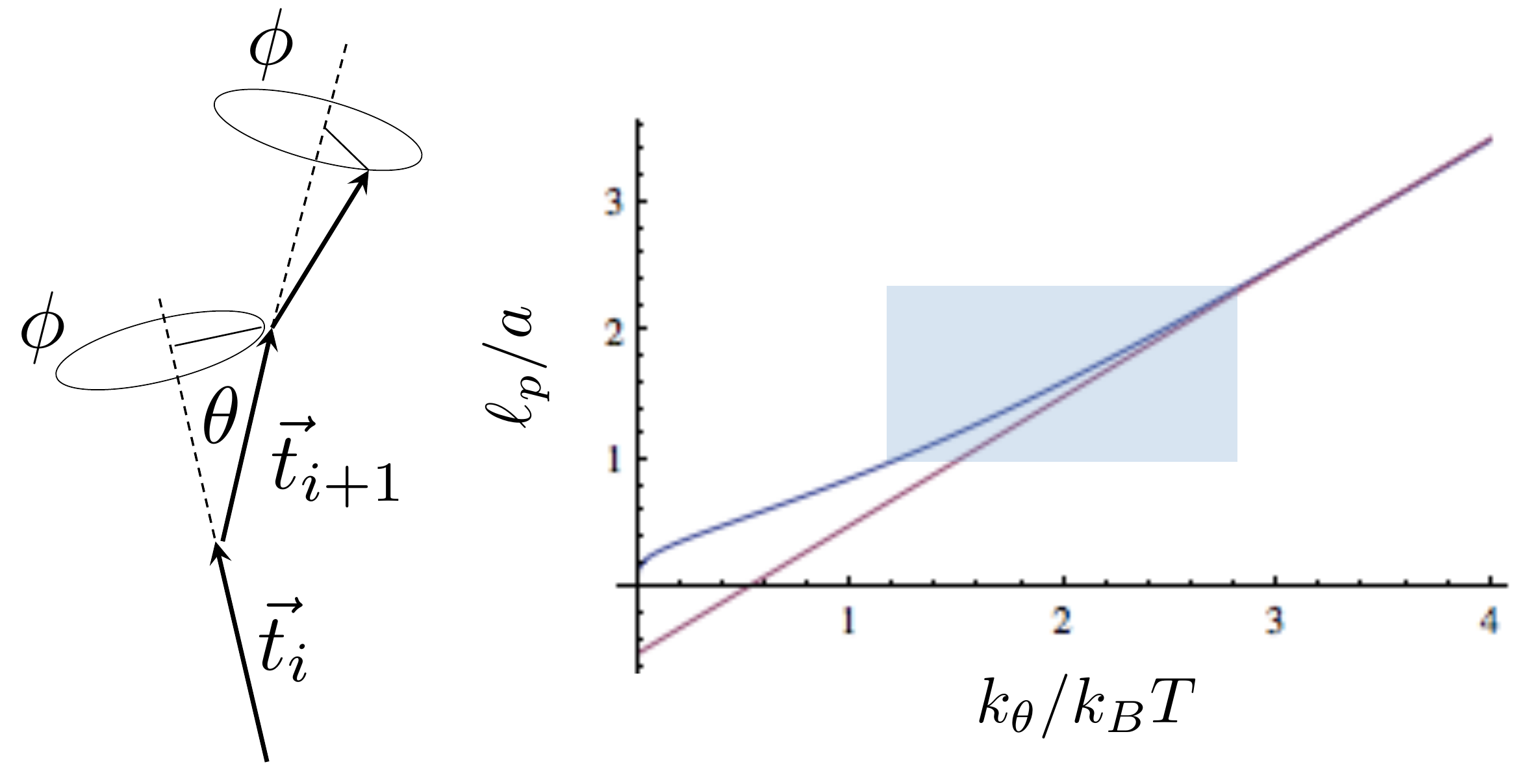}
\caption{Left: section of a freely rotating chain. Right: plot of the function~\eqref{e:lpbc} with 
Eq.~\eqref{e:fa} (blue). The 
purple line is a plot of the stiff polymer approximation $\ell_p  = a (k_\theta/k_BT -1/2)$. The transparent 
rectangle identifies the region $\ell_p/a\in[1,2.5]$ and the corresponding values 
of $\alpha = k_\theta/k_BT$.}
\label{FIG.pol}
\end{figure}

\noindent The bond-bond correlation function (BBCF) for a freely rotating chain is an exponentially
decaying function of the monomer-monomer separation along the 
chain~\cite{polymer}, 
\begin{equation}
\label{e:bbc}
\langle \vec{t}_{i+m} \cdot \vec{t}_{i} \rangle = a^2 \left( 
                                                         \cos \theta 
                                                      \right)^m = a^2 e^{-ma/\ell_p}
\end{equation}
where $\ell_p$ is the chain persistence length, 
\begin{equation}
\label{e:lpfrc}
\ell_p = -\frac{a}{\log (\cos \theta)}
\end{equation}
In eq.~\eqref{e:bbc}, the average is taken over the free rotations about the segment axes 
(free torsions) and $\theta$ is the angle formed by two successive
segments, $\cos \theta = a^{-2} \, \vec{t}_{i+1} \cdot \vec{t}_{i} $, $|\vec{t}_i| = a$
(see Fig.~\ref{FIG.pol}).\\
\indent If the angle $\theta$ is no longer fixed but is controlled by a potential 
energy $V(\theta)$, in the hypothesis that  free torsions along $\phi$ and bending are 
uncoupled degrees of freedom along the chain, the BBCF is still an exponentially
decaying function and the persistence length can be computed as
\begin{equation}
\label{e:lpbc}
\ell_p = -\frac{a}{\log \,\langle \cos \theta \rangle}
\end{equation}
where 
\begin{equation}
\label{e:avcosth}
\langle \cos \theta \rangle = \frac{\int_0^\pi \cos \theta \sin \theta \, e^{-V(\theta)/k_BT}}
                                   {\int_0^\pi \sin \theta \, e^{-V(\theta)/k_BT}}
\end{equation}
%
%
The calculation can be carried out explicitly for example in the case of the cosine angle potential,
$V(\theta) = k_\theta (1 - \cos\theta)$, which reduces to an harmonic bending potential 
for a semiflexible polymer. From eq.~\eqref{e:avcosth}, one gets
\begin{equation}
\label{e:fa}
\langle \cos \theta \rangle = \frac{\alpha-1+(\alpha+1)e^{-2\alpha}}{\alpha(1-e^{-2\alpha})}
\end{equation}
where $\alpha = k_\theta/k_BT$. 
For PEG the persistence length is 
$\ell_p = 3.8$ \AA \ while the monomer length is $a = 1.5$ \AA~\cite{params}.
In our coarse-grained representation the monomer size is 3.5 \AA, which would 
set a lower bound on the angle bending energy $k_\theta \simeq k_BT$.
In our simulations we fixed $k_\theta =  1.8 \, k_BT$, which we used as the 
bending coefficient for the linker (see Fig.~\ref{FIG.pol}, right panel).



\begin{figure*}
\centering
\includegraphics[width=\columnwidth]{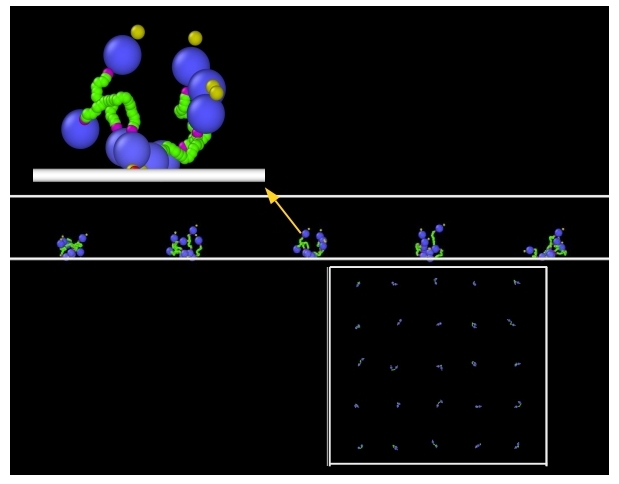}
\caption{\textbf{Simulation for flight/residence time calculation:} Tethered 20-mer SPH diabodies in a $5\times 5$ 
lattice. A similar lattice arrangement was used for all other systems (SBCG systems and SBCG systems with all 
other linker lengths). The distance between neighboring diabodies was increased as the linker 
length increased to avoid any interaction between neighbors.}
\label{FIG.diffsys}
\end{figure*}

\begin{figure*}
\centering
\includegraphics[width=\columnwidth]{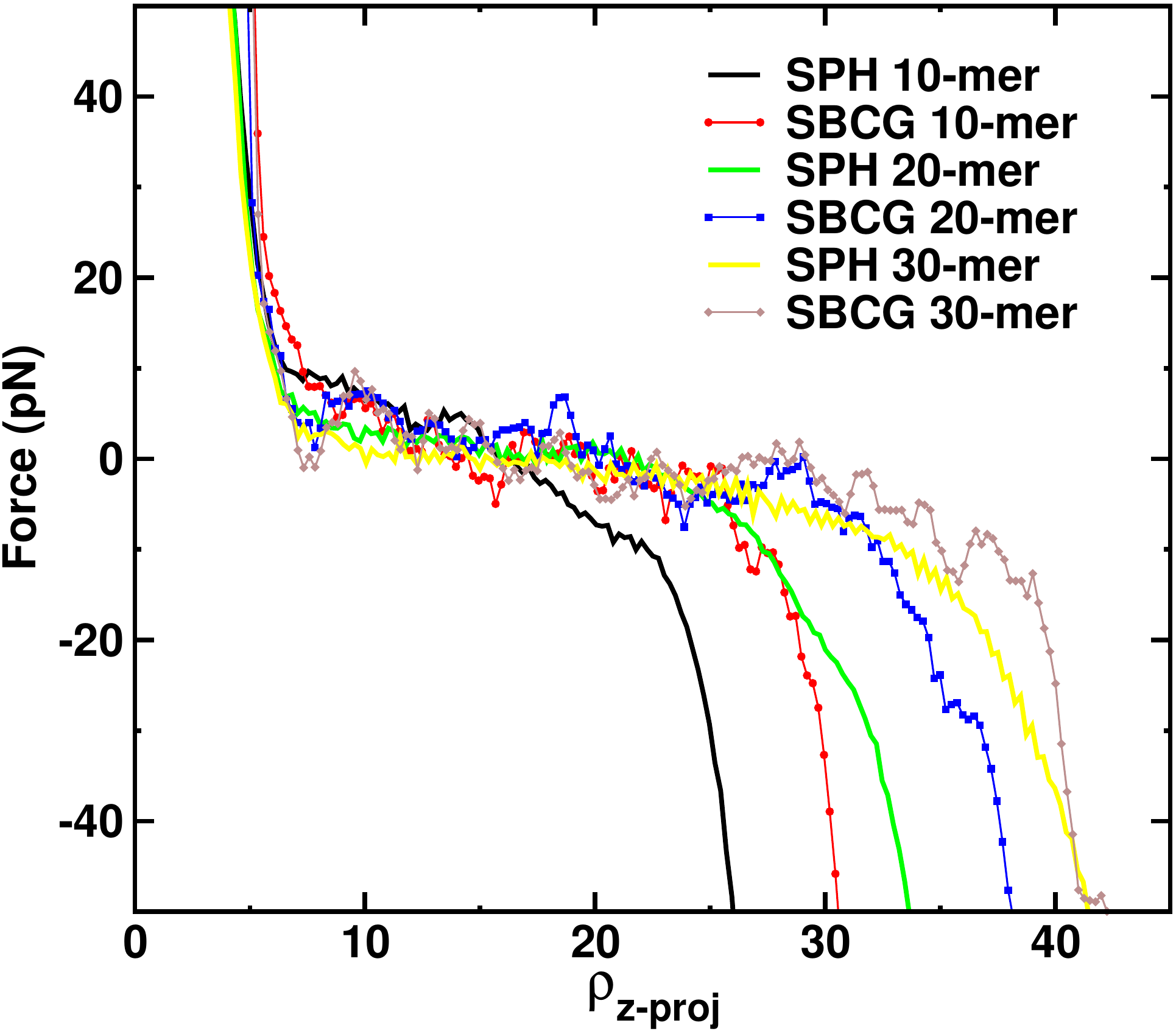}
\caption{\textbf{Force}: Derivative of the PMF profiles as a function of 
$\rho_{z-proj}$ leading to the force on nbd-1 for a given value of the RC. 
The SBCG system experiences larger forces near the wall.}
\label{FIG.force-zproj}
\end{figure*}

\begin{figure*}
\centering
\includegraphics[width=2\columnwidth]{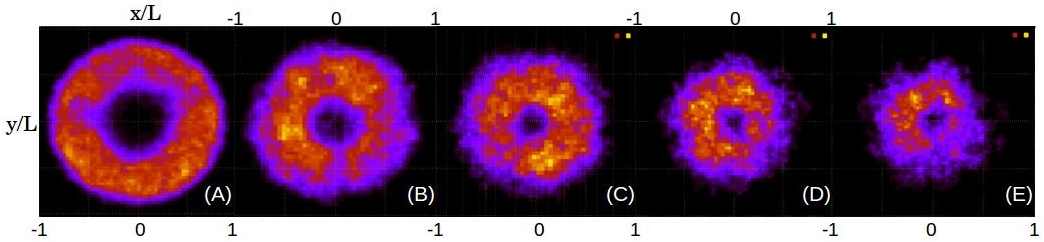}
\caption{The $xy$ distribution of the paratope position averaged over 25 diabodies 
with the condition that the paratope be within a distance of 5 $\sigma$ from 
the lower wall of the box for linker lengths of (A) 10, (B) 20, (C) 30, (D) 40 and (E) 50 monomers.}
\label{FIG.xydist}
\end{figure*}

\begin{figure*}
\centering
\includegraphics[width=\columnwidth]{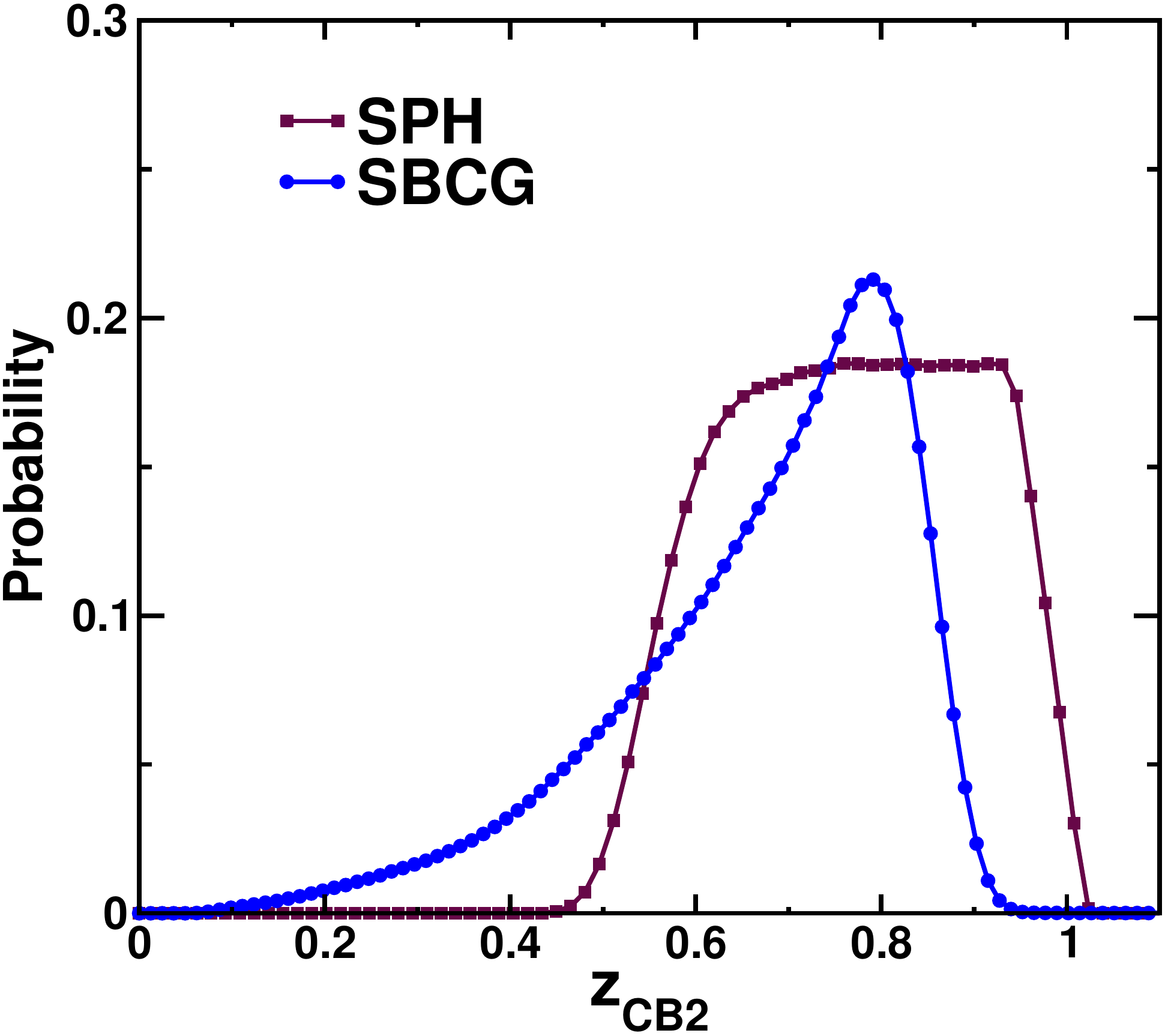}
\caption{Distribution of the normalized height of CB2 from the tethering wall for the 
10-mer SPH system (repulsive wall, $s_{wall}$ = 4.5 for nbd-2) and the 10-mer SBCG system.}
\label{FIG.steric}
\end{figure*}

\begin{figure*}
\centering
\includegraphics[width=2\columnwidth]{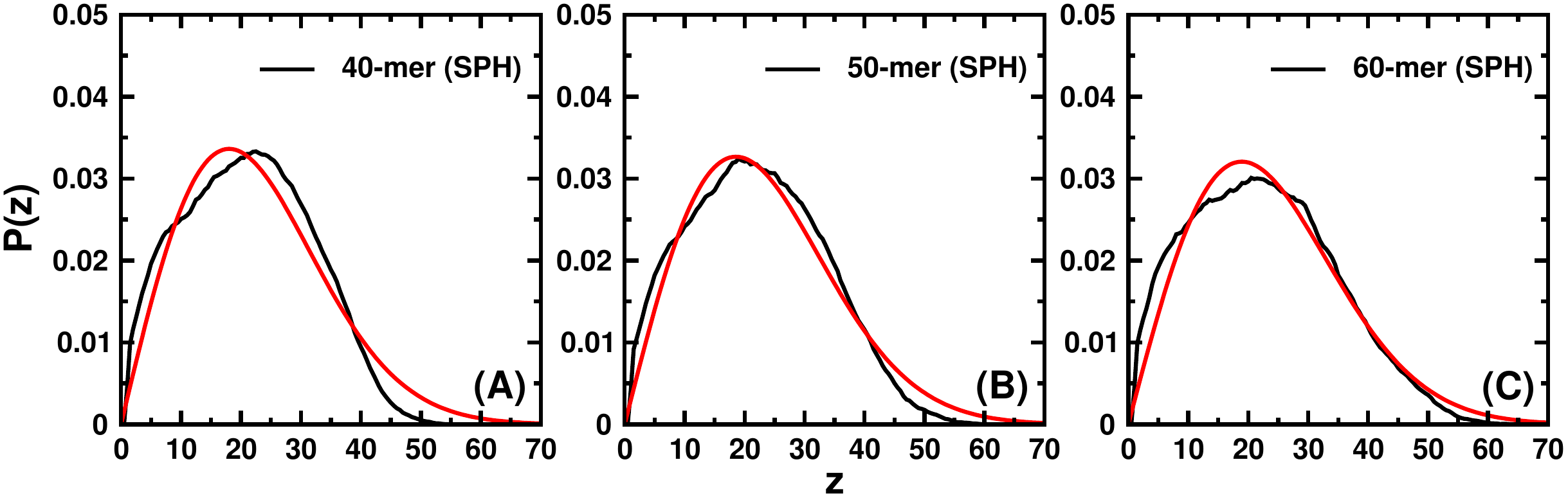}
\caption{Distribution of the $z$-coordinate of P1 for (A) 40-, (B) 50- and 
(C) 60-mer systems along with the fits made using Eq.~(8) 
in the main text with $N_k$ as the free parameter.}
\label{FIG.fit}
\end{figure*}

\begin{figure*}
\centering
\includegraphics[width=0.8\linewidth]{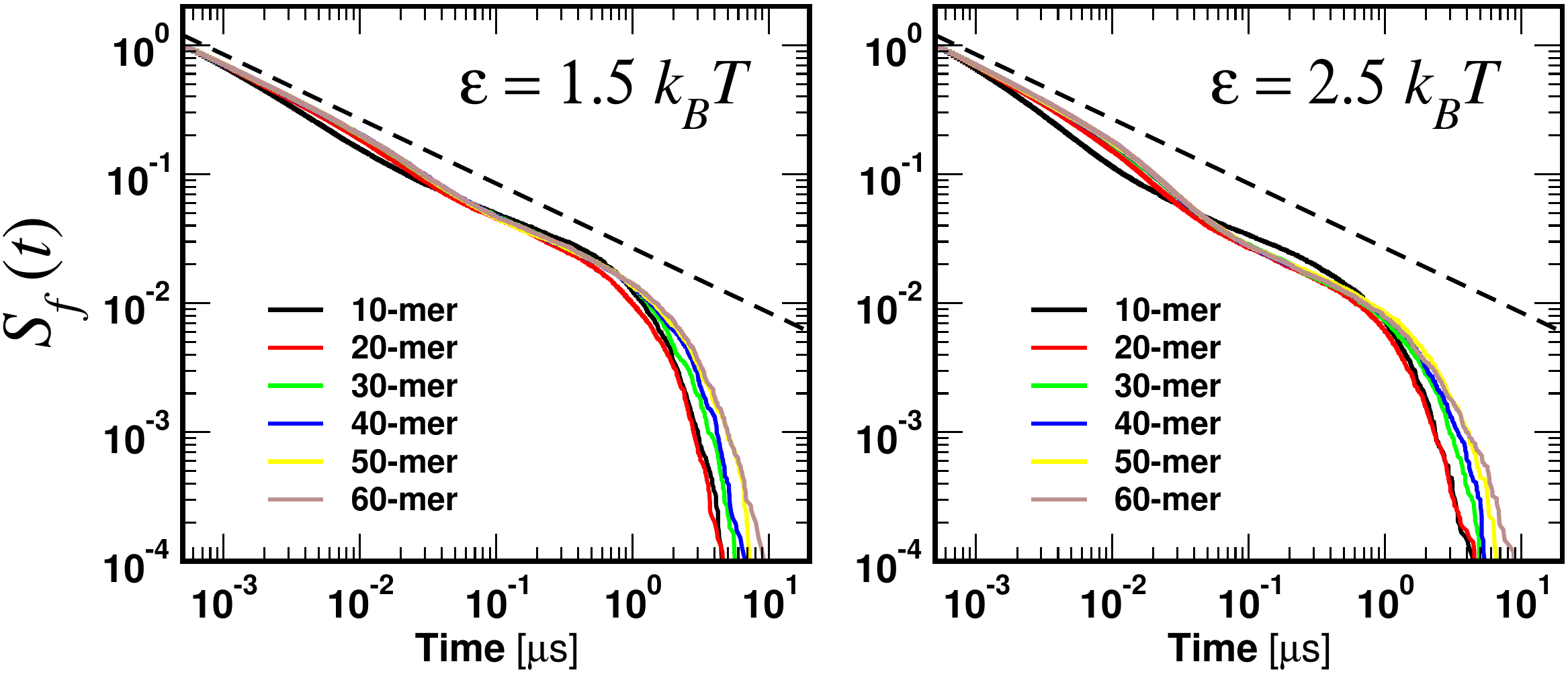}
\includegraphics[width=0.8\linewidth]{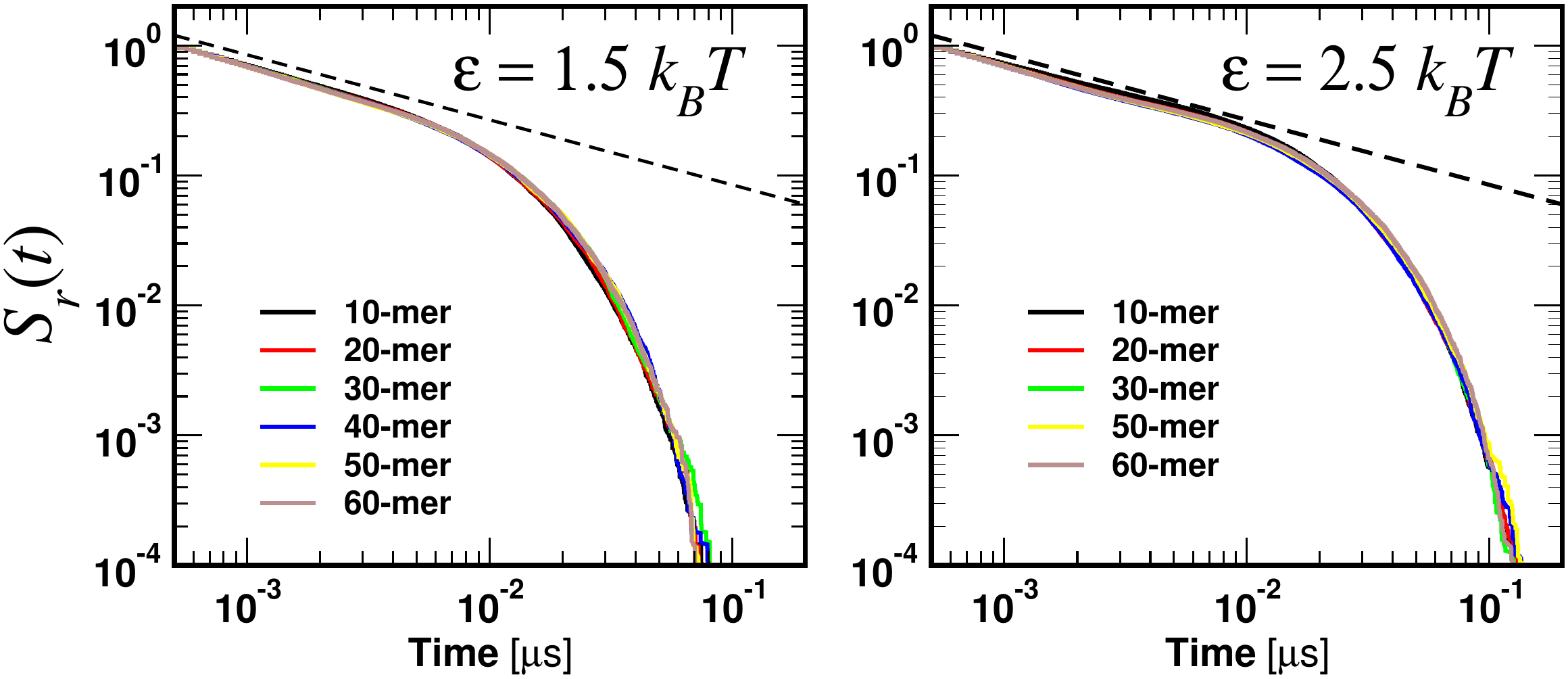}
\caption{(Upper panel) Survival probability of the epitope in the flight 
($z \geq z_{th}$) domain for the SPH (left) system and (right) SBCG system for 
different linker lengths. (Lower panel) Survival probability of the epitope in the 
residence ($z < z_{th}$) domain for the SPH (left) system and (right) SBCG system 
for different linker lengths. The dashed lines are plots of a power law of the kind $t^{-1/2}$. }
\label{FIG.xydist}
\end{figure*}


\begin{figure*}
\centering
\includegraphics[width=0.5\columnwidth]{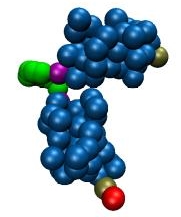}
\caption{Steric repulsion between nbd-1 and nbd-2 for the 10-mer linker
preventing nbd-1 from reaching close to the tethering wall.}
\label{FIG.cb2}
\end{figure*}

\begin{figure*}
\centering
\includegraphics[width=\columnwidth]{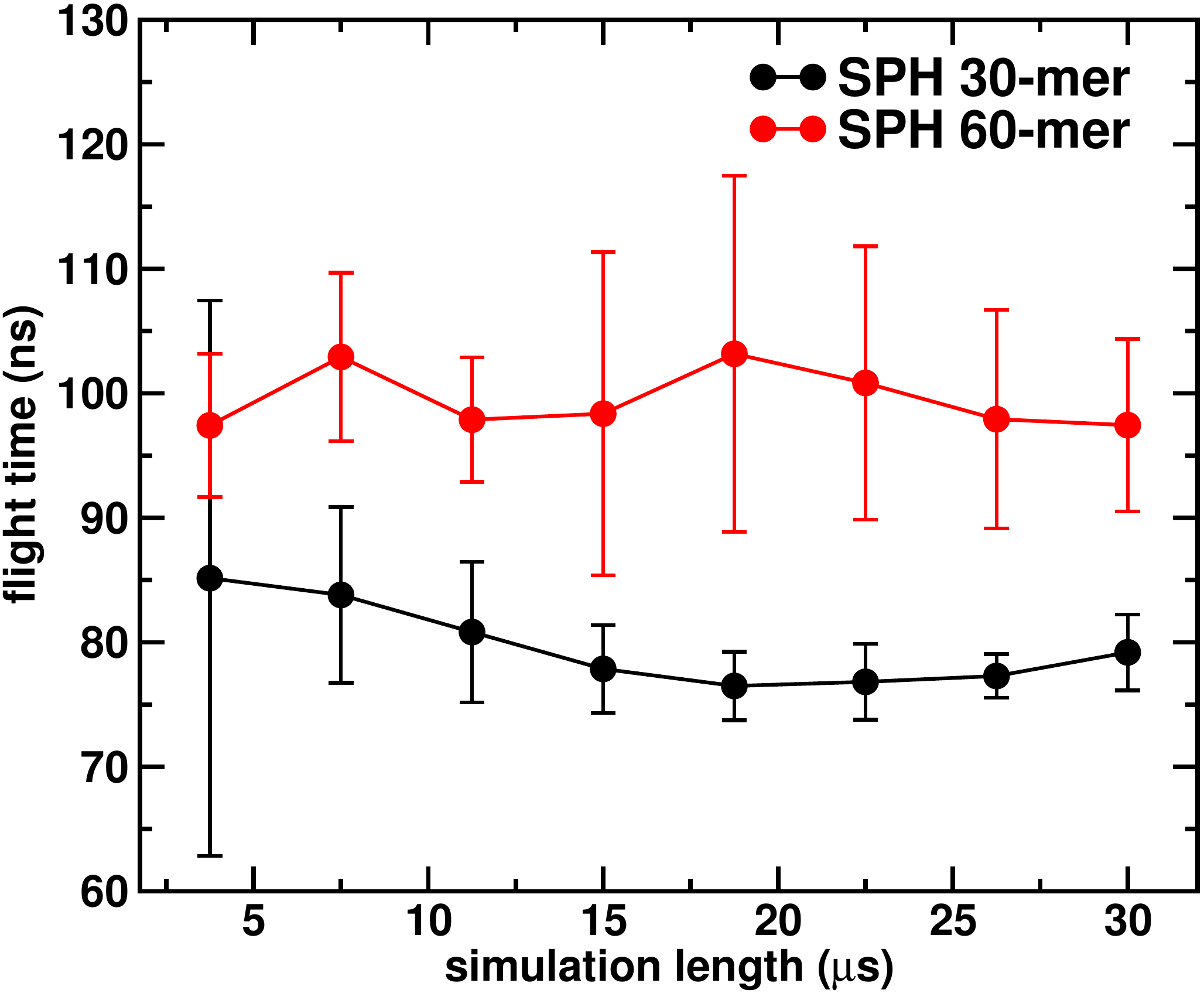}
\caption{Flight time as a function of simulation length.}
\label{FIG.conv}
\end{figure*}

\begin{table*}[!h]
\begin{center}
\caption{\textbf{Flight/Residence times} SPH system, repulsive wall, 
$s_{wall}$ for nbd-2 = 4.5} \label{tab1}
\begin{tabular}{ccc}\hline
Linker length & Flight time (ns) & Residence time (ns)\\
\\ \hline
10& 69.50 $\pm$ 1.92 & 3.24 $\pm$ 0.04\\
20& 68.88 $\pm$ 1.10 & 3.37 $\pm$ 0.08\\
30& 79.18 $\pm$ 3.05 & 3.39 $\pm$ 0.04\\
40& 88.26 $\pm$ 2.22  & 3.45 $\pm$ 0.04\\
50& 89.38 $\pm$ 6.45 & 3.42 $\pm$ 0.05\\
60& 97.45 $\pm$ 6.90 & 3.47 $\pm$ 0.05\\
\hline
\\
\end{tabular}
\end{center}
\end{table*}

\begin{table*}[!bht]
\begin{center}
\caption{\textbf{Flight/Residence times} SPH system, repulsive wall, 
$s_{wall}$ for nbd-2 = 3.0} \label{tab1}
\begin{tabular}{ccc}\hline
Linker length & Flight time (ns) & Residence time (ns)\\
\\ \hline
10 & 59.90 $\pm$ 0.67 & 2.93 $\pm$ 0.07\\
20 & 62.96 $\pm$ 0.75 & 3.21 $\pm$ 0.02\\
30 & 73.83 $\pm$ 2.50 & 3.34 $\pm$ 0.04\\
40 & 82.66 $\pm$ 3.38 & 3.44 $\pm$ 0.04\\
50 & 88.20 $\pm$ 6.12 & 3.38 $\pm$ 0.08\\
60 & 94.89 $\pm$ 5.30 & 3.49 $\pm$ 0.04\\
\hline
\\
\end{tabular}
\end{center}
\end{table*}

\begin{table*}[!bht]
\begin{center}
\caption{\textbf{Flight/Residence times} SPH system, 
attractive wall, $\epsilon = 1.5 k_BT$, $s_{wall}$ for nbd-2 = 4.5} \label{tab1}
\begin{tabular}{ccc}\hline
Linker length & Flight time (ns) & Residence time (ns)\\
\\ \hline
10 & 42.00 $\pm$ 1.53 & 4.76 $\pm$ 0.06\\
20 & 37.20 $\pm$ 1.67 & 4.76 $\pm$ 0.03\\
30 & 45.50 $\pm$ 1.61 & 4.83 $\pm$ 0.05\\
40 & 48.31 $\pm$ 1.24 & 4.81 $\pm$ 0.06\\
50 & 51.15 $\pm$ 0.92 & 4.82 $\pm$ 0.06\\
60 & 53.84 $\pm$ 5.89 & 4.85 $\pm$ 0.04\\
\hline
\\
\end{tabular}
\end{center}
\end{table*}

\begin{table*}[!bht]
\begin{center}
\caption{\textbf{Flight/Residence times} SPH system, attractive wall, 
$\epsilon = 2.5 k_BT$, $s_{wall}$ for nbd-2 = 4.5} \label{tab1}
\begin{tabular}{ccc}\hline
Linker length & Flight time (ns) & Residence time (ns)\\
\\ \hline
10 & 26.27 $\pm$ 0.70 & 7.52 $\pm$ 0.06\\
20 & 23.27 $\pm$ 0.45 & 7.10 $\pm$ 0.06\\
30 & 27.89 $\pm$ 1.39 & 7.10 $\pm$ 0.11\\
40 & 29.54 $\pm$ 1.56 & 6.90 $\pm$ 0.03\\
50 & 32.71 $\pm$ 1.67 & 7.14 $\pm$ 0.14\\
60 & 32.06 $\pm$ 2.50 & 7.30 $\pm$ 0.13\\
\hline
\\
\end{tabular}
\end{center}
\end{table*}

\begin{table*}[!bht]
\begin{center}
\caption{\textbf{Flight/Residence times} SBCG system, repulsive wall} \label{tab1}
\begin{tabular}{ccc}\hline
Linker length & Flight time (ns) & Residence time (ns)\\
\\ \hline
10 & 53.72 & 1.90 \\
20 & 45.03 & 1.83 \\
30 & 49.50 & 1.86 \\
\hline
\\
\end{tabular}
\end{center}
\end{table*}

\end{document}